\documentclass[prl, twocolumn,superscriptaddress,amsmath,amssymb,showpacs,floatfix,preprintnumbers]{revtex4-2}

\usepackage{amsmath}
\usepackage{graphicx}
\usepackage{dcolumn}
\usepackage{xcolor}
\usepackage{physics}
\usepackage{xr}
\usepackage{siunitx}
\usepackage{tabularx}
\usepackage{makecell}
\usepackage{multirow}
\DeclareGraphicsExtensions{.png .jpg .pdf}

\usepackage[normalem]{ulem}

\makeatletter
\newcommand*{\addFileDependency}[1]{
  \typeout{(#1)}
  \@addtofilelist{#1}
  \IfFileExists{#1}{}{\typeout{No file #1.}}
}
\makeatother
 
\newcommand*{\myexternaldocument}[1]{%
    \externaldocument{#1}%
    \addFileDependency{#1.tex}%
    \addFileDependency{#1.aux}%
}

\myexternaldocument{si}

\begin{document}

\title{Moir\'e polar vortex, flat bands and Lieb lattice in twisted bilayer BaTiO$_3$}

\author{Seungjun Lee}
\email{seunglee@umn.edu}
\affiliation{Department of Electrical and Computer Engineering, University of Minnesota, Minneapolis, Minnesota 55455, USA}

\author{D. J. P. de Sousa}
\email{sousa020@umn.edu}
\affiliation{Department of Electrical and Computer Engineering, University of Minnesota, Minneapolis, Minnesota 55455, USA}

\author{Bharat Jalan}
\affiliation{Department of Chemical Engineering and Materials Science, University of Minnesota, Minneapolis, Minnesota 55455,USA}

\author{Tony Low}\email{tlow@umn.edu}
\affiliation{Department of Electrical and Computer Engineering, University of Minnesota, Minneapolis, Minnesota 55455, USA}
\affiliation{Department of Physics, University of Minnesota, Minneapolis, Minnesota 55455, USA}

\begin{abstract}
Advances in material fabrication techniques and growth methods have opened up a new chapter for twistronics, in the form of twisted freestanding three-dimensional material membranes. Through first-principles calculations based on density functional theory, we investigate the crystal and electronic structures of twisted bilayer BaTiO$_3$. Our findings reveal that large stacking fault energy leads to chiral in-plane vortex pattern that was recently observed in experiments. 
Moreover, we also found non-zero out-of-plane local dipole moments, indicating that the strong interlayer interaction might offer promising strategy to stabilize ferroelectric order in the two-dimensional limit.
Remarkably, the vortex pattern in the twisted BaTiO$_3$ bilayer support localized electronic states with quasi-flat bands, associated with the interlayer hybridization of oxygen $p_z$ orbitals. We found that the associated band width reaches a minimum at $\sim$19$^{\circ}$ twisting, configuring the largest magic angle in moir\'e systems reported so far.
Further, the moir\'e vortex pattern bears a striking resemblance to two interpenetrating Lieb lattices and corresponding tight-binding model provides a comprehensive description of the evolution the moir\'e bands with twist angle and reveals the topological nature of these states.

\end{abstract}

\maketitle

\section{Introduction}

The ability to precisely control the twist angle between van der Waals (vdW) materials has opened up an emerging field of research, called '`twistronics''.~\cite{carr2020electronic,andrei2020graphene}
The first major breakthrough arose from the discovery of magic-angle twisted bilayer graphene,~\cite{bistritzer2011moire,cao2018correlated,cao2018unconventional} followed by other vdW materials,~\cite{wu2018hubbard,naik2018ultraflatbands,xian2019multiflat} and further open up tremendous opportunities in the exploration of new emergent phenomena. Advances in material fabrication technologies have further expanded the possibilities, enabling the design of entirely new types of moir\'e materials. 
Utilizing vdW materials~\cite{kim2017remote,kim2022remote,manzo2022pinhole,jang2023thru,yoon2022freestanding} or a sacrificial layer~\cite{lu2016synthesis,hong2017two,ji2019freestanding,varshney2024hybrid} , it is now possible to realize atomically thin and freestanding membrane consisting of three-dimensional (3D) oxide materials.
Recently, moir\'e perovskites bilayer membranes were realized experimentally.~\cite{shen2022observation,li2022stacking,sanchez20242d}
Due to the interlayer interaction in these moir\'e oxides, unique vortex patterns that possess strong local polarization were also observed,~\cite{sanchez20242d} a signature of the broken symmetry due to the chiral interlayer coupling. These developments thus ushered in a new chapter in oxide twistronics.

Despite these first experimental observations,~\cite{sanchez20242d} little is known about the origin and character of the polar vortex, the Moire electronic structures, the magic angle and associated flat band physics in this material. 
In this work, we investigated the crystal and electronic structure of twisted 2L BaTiO$_3$, by means of first-principles calculations based on density functional theory (DFT)~\cite{Kohn1965,Kresse1996}.
Our findings reveal a substantial generalized stacking fault energy (GSFE) in 2L BaTiO$_3$ with a BaO-BaO type interface. This results in significant intralayer relaxation and induces chiral in-plane moir\'e vortex patterns at finite twist angles $\theta_t$, in full agreement with recent experimental observations~\cite{sanchez20242d}.
Notably, the chirality of the vortex patterns are opposing for the two layers. 
We also find that interlayer interactions give rise to non-zero local out-of-plane dipole moments and strong electrostatic surface potentials. 
Thus, twisted 2L perovskites can serve as ``smart'' substrates for epitaxial growth, or to imprint electrostatic superlattices on vdW materials.

Pertaining to the electronic structures, our results indicate that the moir\'e vortex pattern hosts localized electrons states with quasi-flat bands over a wide range of twist angles. We found that the maximum band flatness occurs at ${\theta_t}{\approx}19^{\circ}$, configuring the largest magic angle reported in literature, to the best of our knowledge. Remarkably, the low energy bands formed from these localized states are well described by a tight-binding model on two interpenetrating Lieb lattices, whose sites correspond to the distinct local vortex patterns. Thus, moir\'e states in twisted 2L BaTiO$_3$ hold the potential to enable the first realization of an inorganic electronic Lieb lattice.

\section{Results and discussion}
\subsection{Crystal and electronic structures of untwisted bilayer BaTiO$_3$}

\begin{figure}[h]
\includegraphics[width=1.00\columnwidth]{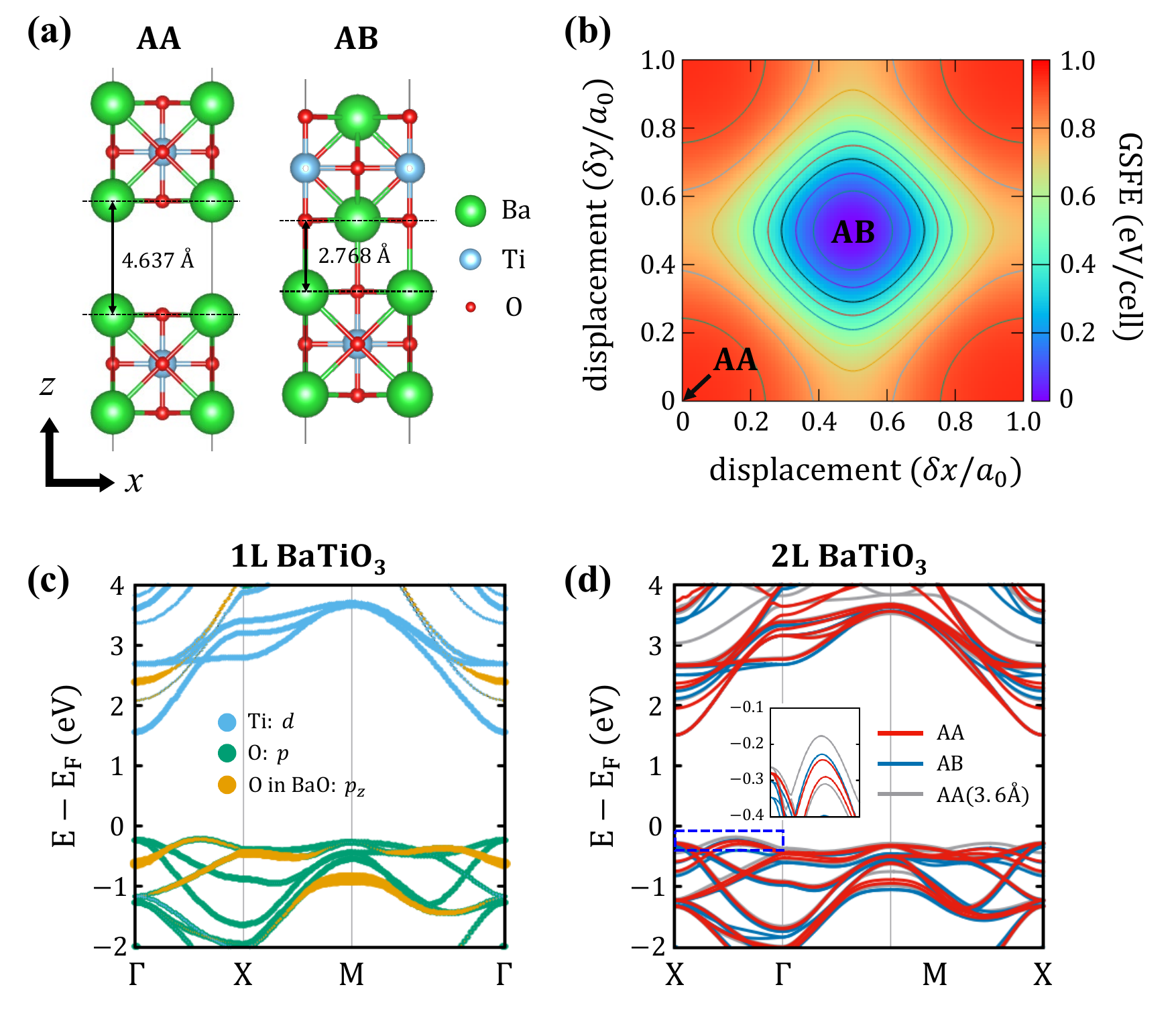}
\caption{\textbf{Giant stacking energy fault and electronic structures of untwisted bilayer BaTiO$_3$}
(a) Crystal structures of bilayer (2L) BaTiO$_3$ with AA and AB stacking configurations, and (b) generalized stacking fault energy (GSFE) map for 2L BaTiO$_3$. (c) Orbital-resolved electronic structure of monolayer BaTiO$_3$. (d) Electronic structures of 2L BaTiO$_3$ with AA, AB, AA$^{\prime}$ stacking configurations. AA(3.6~{\AA}) stacking has the same in-plane atomic arrangement to AA stacking, but has 1~{\AA} smaller interlayer distance compared to the AA stacking. Inset visualizes the band structure near the valence band maximum.
}\label{fig1}
\end{figure}

In the ABO$_3$ perovskites, there are two possible types of interfaces, related to the AO and BO$_2$ surfaces. Therefore, in twisted 2L BaTiO$_3$, we have a total of three different interface configurations, namely BaO-BaO, TiO$_2$-TiO$_2$, and BaO-TiO$_2$. Our DFT calculations indicate that both TiO$_2$-TiO$_2$ and BaO-TiO$_2$ interfaces give rise to complex and unsaturated chemical bonds between the interfaces, resulting in many defect-like states in their electronic structures. (See Figs~\ref{figs_Tio2} and~\ref{figs_Bao_Tio2} in Supplementary Information (SI))
On the other hand, BaO-BaO interface has vdW-like interlayer interaction. 
On the experimental front, various surface termination have been reported in ABO$_3$ perovskites.~\cite{ricciardulli2021emerging}
Among them, one recent experimental study observed the topmost BaO surface on the BaTiO$_3$ thin film, suggesting that the fabrication of BaO-BaO interfaces is possible in experiments.~\cite{deleuze2022nature}
Since membranes mediated by vdW-like forces are more amenable to “twistronics” approaches, we shall focus exclusively on membranes with BaO termination.

We begin by discussing the untwisted bilayer structures. With BaO-BaO surface termination, the chemical formula of a BaTiO$_3$ membrane can be described as Ba$_{n+1}$Ti$_n$O$_{3n+1}$ with $n$ being infinity.
As a first step, we performed DFT calculations for $n$=1 case corresponding to 2 layers of BaTiO$_3$ sandwiching one layer of BaO and made two different stacking configurations, AA and AB, as shown in Fig.~\ref{fig1}(a). For more detailed description of DFT calculations, see Method section.
Due to the opposite charge states in Ba and O, the interlayer distance in the AA configuration is much larger than AB. To understand the stacking energy landscape, we calculated generalized stacking fault energy (GSFE)~\cite{zhou2015van,carr2018relaxation} as function of the in-plane sliding, as shown in Fig.~\ref{fig1}(b). 
The GSFE corresponding to the AA stacking is almost 1~eV, which is orders of magnitude larger than that of typical vdW materials, such as graphene (20~meV/cell) and MoS$_2$ (60~meV/cell).~\cite{carr2018relaxation}
On the other hand, some layered materials exhibit comparable GSFE to 2L BaTiO$_3$, such as SnS/SnSe heterostructure (0.88~eV/cell).~\cite{ozccelik2018tin}
It means that the interlayer bonding characteristics of 2L BaTiO$_3$ is in the range of vdW-like rather than covalent and ionic.
Hence, substantial atomic relaxation is expected to take place in the twisted configurations, which will be addressed in upcoming sections. 

Figure~\ref{fig1}(c) shows the orbital-resolved electronic structure of monolayer (1L) BaTiO$_3$. Its bandgap is calculated to be 1.78~eV within PBE XC functional, which is slightly larger than that of bulk BaTiO$_3$ (1.62~eV) calculated by the same computational method. While the character of the parabolic conduction band is mainly due to Ti $d$ orbitals, O $p$ orbitals dominates the valence band. Also, we can find sizable contributions of $p_z$ states of O in BaO plane in valence band maximum (VBM, along the $\Gamma - X$ path in the Brillouin zone). 
Figure~\ref{fig1}(d) shows the electronic structures of bilayer BaTiO$_3$ with AA and AB stacking configurations. Their electronic structures are very similar with that of 1L BaTiO$_3$, implying a weak interlayer interaction in both configurations. To understand the role of the O $p_z$ states, we constructed an artificial structure with an AA(3.6~{\AA}) stacking configuration, which has an interlayer distance reduced by 1~{\AA} as compared to the true AA stacking configuration, and {found a higher VBM as shown in grey color in Fig.~\ref{fig1}(d).} 
Thus, it suggest that the interlayer O-O interaction controls the hybridization between the valence bands of the constituents layers. Note that, interlayer coupling does not affect conduction bands due to its Ti $d$ orbital.

\begin{figure*}[t]
\includegraphics[width=2.00\columnwidth]{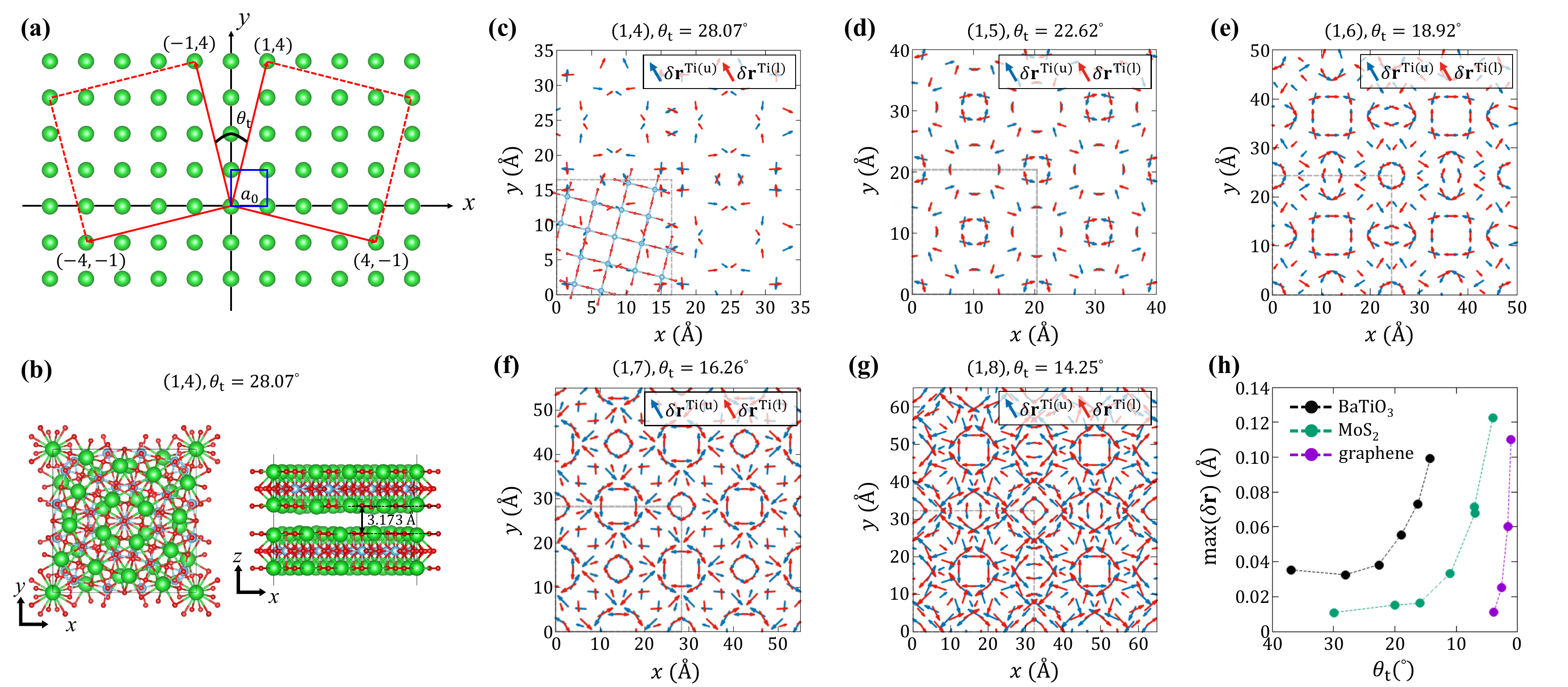}
\caption{\textbf{In-plane polar vortex in twisted bilayer BaTiO$_3$} (a) Schematics of the supercell construction method with non-zero twist angle (${\theta}_{\rm{t}}$) in a square lattice. Green sphere indicates Ba sublattice and blue square and $a_0$ indicate unitcell and lattice constant of BaTiO$_3$. The size of the supercell and corresponding ${\theta}_{\rm{t}}$ can be determined by four lattice vectors (red arrows), which are ($n$,$m$), ($m$,$-n$), ($-n$,$m$), and ($-m$,$-n$), respectively, where ($n$,$m$) indicate lattice vector defined as $n a_0\hat{x}+m a_0\hat{y}$ and $a_0$ indicates the lattice constant of BaTiO$_3$. Then, ${\theta}_{\rm{t}}$ was evaluated to be ${\theta}_{\rm{t}}=2~\rm{arctan}$($n$/$m$).
(b) Top and side views of the relaxed crystal structure of 2L BaTiO$_3$ with ${\theta}_{\rm{t}}=28.07^{\circ}$. 
The lattice vector of this supercell can be described as (1,4).
(c-g) In-plane Ti displacement vector ($\delta{\mathbf{r}}^{\rm{Ti}}$) maps for supercell structures with ${\theta}_{\rm{t}}=$28.07, 22.62, 18.92, 16.26, and, 14.25$^{\circ}$, respectively. The $\delta{\mathbf{r}}^{\rm{Ti}}$ was amplified by a factor of 50 for easy visualization.
Blue and red arrows indicate $\delta{\mathbf{r}}^{\rm{Ti}}$ of upper (u) and lower (l) layers, and grey dahsed lines show supercell structures. In (c), we overlaid atomic structure of TiO$_2$ sublayer for better understanding. (h) Maximum magnitude of in-plane $\delta{\mathbf{r}}$ of twisted 2L BaTiO$_3$, graphene,~\cite{cantele2020structural} and MoS$_2$~\cite{kim2022anomalous} as function of ${\theta}_{\rm{t}}$.
}\label{fig2}
\end{figure*}

\subsection{In-plane polar vortex in twisted bilayer BaTiO$_3$}

Due to the two-dimensional square lattice structure of 1L BaTiO$_3$, we can systematically construct commensurate supercell structures of 2L BaTiO$_3$ with finite twist angle (${\theta}_{\rm{t}}$) without any additional strain, as illustrated in Fig.~\ref{fig2}(a). 
The size of the supercell and corresponding ${\theta}_{\rm{t}}$ can be determined by four lattice vectors, corresponding to the combinations ($n$,$m$), ($m$,$-n$), ($-n$,$m$), and ($-m$,$-n$), respectively, where ($n$,$m$) indicate lattice vector defined as $n a_0\hat{x}+m a_0\hat{y}$ and $a_0$ indicates the lattice constant of 1L BaTiO$_3$. Then, ${\theta}_{\rm{t}}$ was evaluated to be ${\theta}_{\rm{t}}=2~\rm{arctan}$($n$/$m$).
Note that, when $n$+$m$=$\textrm{even}$, we can generally find a smaller supercell structure with the same ${\theta}_{\rm{t}}$, but characterized by the lattice vector of ($(n+m)/2$,$(m-n)/2$). However, in this work, we also utilize the lattice vector ($n$,$m$) when $n$+$m$=$\textrm{even}$, for consistent visualization purposes.
Using this approach, we constructed a total of five supercell structures with ${\theta}_{\rm{t}}$ ranging from 12.7$^{\circ}$ to 28.07$^{\circ}$.
(See Fig.~\ref{figs_structure} in SI for their relaxed atomic structures)
Figure~\ref{fig2}(b) shows the top and side views of one of twisted 2L BaTiO$_3$ structure with 28.07$^{\circ}$ twist angle. The averaged interlayer distance was 3.173~$\rm{\AA}$, which is intermediate value between AA and AB, indicating a vdW-type interlayer interaction. 
The obtained averaged interlayer distance decreases slowly as ${\theta}_{\rm{t}}$ is decreased. For example, the interlayer distance was calculated to be 3.131~{\AA} for 18.92$^{\circ}$, and 3.122~{\AA} for 14.25$^{\circ}$. Within ${\theta}_{\rm{t}}$ ranging from 28.07$^{\circ}$ to 14.25$^{\circ}$, we cannot observe a large variation of the interlayer distance between AA and AB sites.

To understand the origin of polar vortex pattern observed in the recent experiments~\cite{sanchez20242d}, we calculated in-plane Ti displacement vector ($\delta{\mathbf{r}}^{\rm{Ti}}$) defined as $\delta{\mathbf{r}}^{\rm{Ti}}=\mathbf{r}^{\rm{Ti}}_{\rm{relaxed}}-\mathbf{r}^{\rm{Ti}}_{\rm{ini}}$, where $r^{\rm{Ti}}_{\rm{relaxed}}$ is the relaxed position of Ti and $r^{\rm{Ti}}_{\rm{ini}}$ the initial position with cubic symmetry.
Then, we represented $\delta{\mathbf{r}}^{\rm{Ti}}$ for upper (u) and lower (l) layers, as shown in Fig.~\ref{fig2}(c-g). Our results indicate the emergence of chiral vortex displacement patterns for all twist angles. 
Notably, the vortex patterns are opposite for upper and lower layers. 
In addition, the in-plane displacement is larger for smaller twist angles (large mori\'e structure), as shown in Fig.~\ref{fig2}(h), similar to the twisted 2L graphene and MoS$_2$~\cite{cantele2020structural,kim2022anomalous} systems.
More interestingly, the observed in-plane displacement are much larger than those observed in other 2D materials at similar twist angles, which we attribute to the substantial GSFE.
It is worth noting that the physical origin of our observed polar vortex in twisted perovskites is fundamentally distinct from the recently reported pattern in twisted hexagonal boron nitride.~\cite{bennett2023polar}
The former originates from intralayer relaxation due to the interlayer interaction, while the later originates from the stacking order. 

\subsection{Out-of-plane local dipole moments emerging in twisted bilayer BaTiO$_3$}

\begin{figure}[h]
\includegraphics[width=1.00\columnwidth]{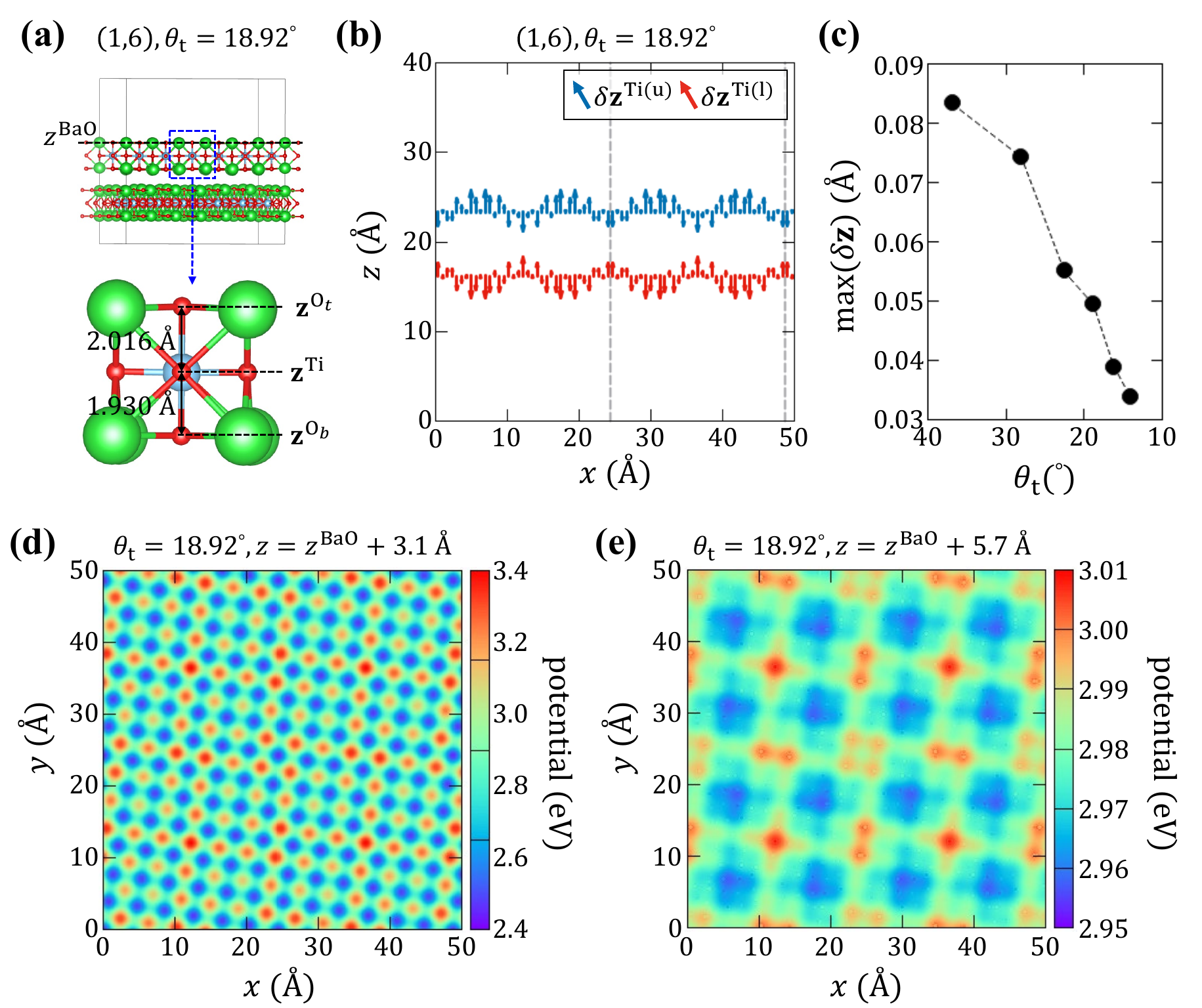}
\caption{\textbf{Out-of-plane local dipole moments in twisted bilayer BaTiO$_3$}
(a) Side view of the relaxed crystal structure of 2L BaTiO$_3$ with ${\theta}_{\rm{t}}=18.92^{\circ}$ and detailed atomic arrangement of at the center of the upper layer. Out-of-plane Ti displacement vector ($\delta{\mathbf{z}^{\rm{Ti}}}$) was defined as $\delta{\mathbf{z}^{\rm{Ti}}}=z^{\rm{Ti}}-(z^{\rm{O_t}}+z^{\rm{O_b}})/2$. 
(b) Calculated $\delta{\mathbf{z}^{\rm{Ti}}}$ of 2L BaTiO$_3$ with ${\theta}_{\rm{t}}=18.92$. 
Blue and red arrows indicate $\delta{\mathbf{z}^{\rm{Ti}}}$ of upper (u) and lower (l) layers, and $\delta{\mathbf{z}^{\rm{Ti}}}$ was amplified by a factor of 50 for easy visualization.
(c) Maximum magnitude of $\delta{\mathbf{z}^{\rm{Ti}}}$ of twisted 2L BaTiO$_3$ as function of ${\theta}_{\rm{t}}$
(d, e) Surface potential profiles of ${\theta}_{\rm{t}}=18.92$ evaluated at distance 3.1~{\AA} and 5.7~{\AA} from the top BaO surface (z$^{\rm{BaO}}$, marked in (a)), respectively. 
}\label{fig3}
\end{figure}

In addition to in-plane polar vortex structure, our first principles calculations also indicate the emergence of local out-of-plane Ti displacements ($\delta{\mathbf{z}^{\rm{Ti}}}$), which can generate non-zero local dipole moment. 
Here, we defined $\delta{\mathbf{z}^{\rm{Ti}}}$ using the z positions of two O atoms which are placed on top ($\rm{O_t}$) and bottom ($\rm{O_b}$) of Ti, described as $\delta{\mathbf{z}}^{\rm{Ti}}=z^{\rm{Ti}}-(z^{\rm{O_t}}+z^{\rm{O_b}})/2$, as shown in Fig.~\ref{fig3}(a).
Then, we  calculated $\delta{\mathbf{z}}$ for ${\theta}_{\rm{t}}=$18.92$^{\circ}$ structure, as shown in Fig.~\ref{fig2}(b). We found non-zero $\delta{\mathbf{z}}$ in twisted 2L BaTiO$_3$, implying emergence of non-zero out-of-plane dipole moment even in 1L thickness. 
We have also calculated $\delta{\mathbf{z}}^{\rm{Ti}}$ for all other structures, which is presented in Fig.~\ref{fig3}(c) along with their maximum values as function of ${\theta}_{\rm{t}}$. Besides non-zero $\delta{\mathbf{z}}^{\rm{Ti}}$ for all structures, we found a negative correlation between  $\delta{\mathbf{z}}^{\rm{Ti}}$ and ${\theta}_{\rm{t}}$.
In perovskites thin film, out-of-plane dipole moment is strongly suppressed below a critical thickness due to the depolarization fields from the dangling bonds at the surface~\cite{junquera2003critical,kim2005critical}.
Our results imply that the moir\'e interlayer interaction might offer a promising strategy to stabilize ferroelectric order in the two-dimensional limit.

We further analyze the out-of-plane displacement through visualizing the surface potential variations of ${\theta}_{\rm{t}}=18.92$ evaluated at distance 3.1~{\AA} and 5.7~{\AA} from the top BaO surface (z$^{\rm{BaO}}$, marked in Fig.~\ref{fig3}(a)), as shown in Fig.~\ref{fig3}(d) and (e), respectively. 
At relatively short distances (3.1~{\AA}), a strong atomic potential difference between Ba and O directly maps into surface potential distribution. 
As a results, higher (lower) potential, red (blue) color, directly describes position of O (Ba) atoms, as shown in Fig.~\ref{fig3}(e). 
At larger distances (Fig.~\ref{fig2}(h)), an underlying long-range order with sizable potential differences of up to 60~meV can be clearly visualized (See Fig.~\ref{figs_zz_detail} in SI for more details).
We note that this long-range order resembles the in-plane polar vortex pattern shown in Fig.~\ref{fig3}(f). In addition, the higher potential magnitudes (red) correspond to the local regions where the out-of-plane displacements $\delta{\mathbf{z}^{\rm{Ti}}}$ point inward, represented in Fig.~\ref{fig2}(f), suggesting that the local out-of-plane dipole moments play a significant role in generating surface potential map. We also found a similar long-range order in other twist angles, as shown in Fig.~\ref{figs_zz_others} in SI.
It was recently reported that potential variations emerging from moir\'{e} pattern can be utilize an effective substrate that can modulate physical properties of two-dimensional materials.~\cite{kim2024electrostatic}
Therefore, the twisted BaTiO$_3$ can be a promising substrate material to manipulate the physical properties of adjacent functional layers.


\subsection{Topological Lieb lattice in twisted bilayer BaTiO$_3$}

\begin{figure}[t]
\includegraphics[width=1.00\columnwidth]{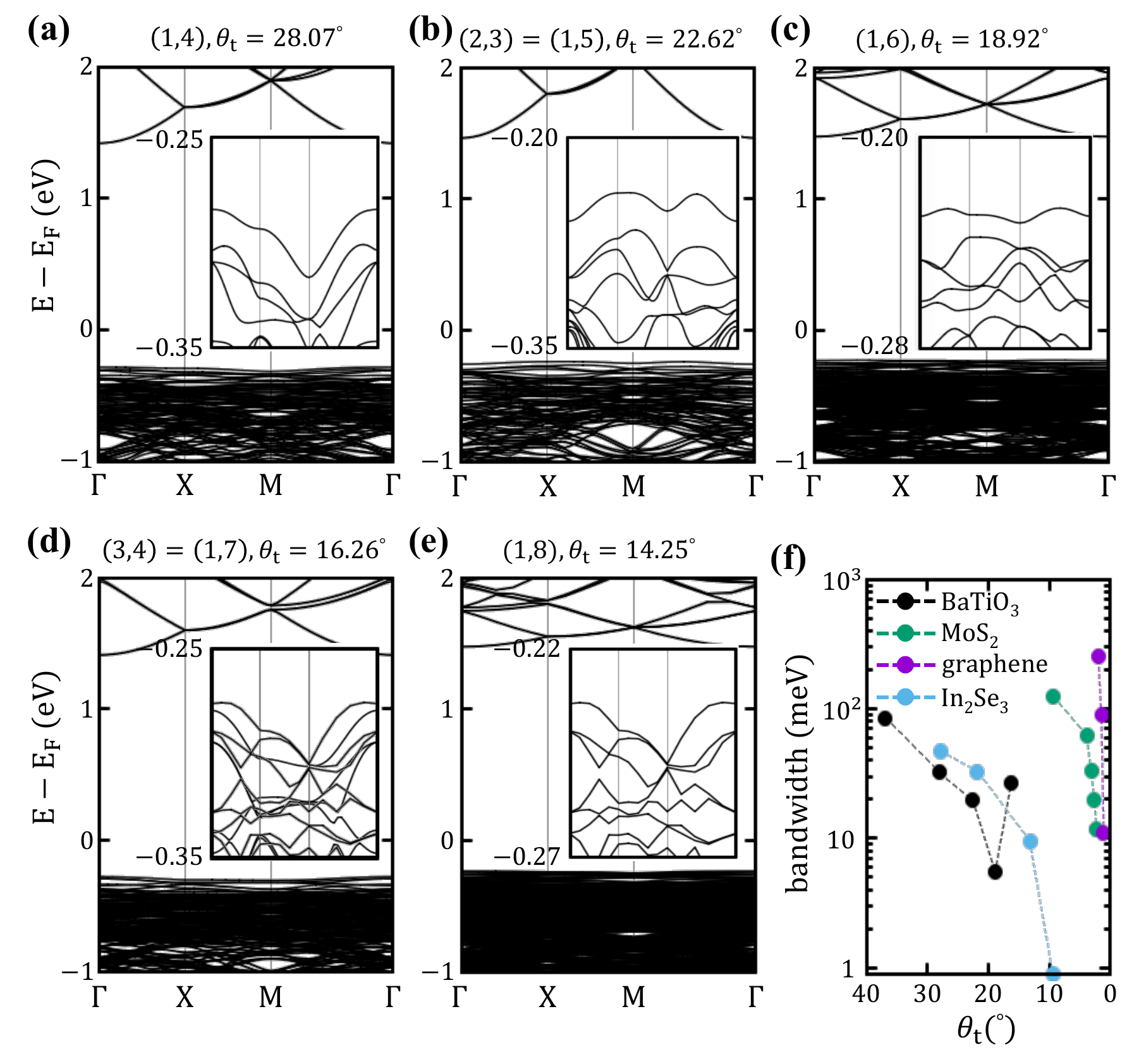}
\caption{\textbf{Mori\'e flat bands in twisted bilayer BaTiO$_3$}
Electronic structure of twisted 2L BaTiO$_3$. (a-e) Evolution of the moir\'e bands for 28.07$^{\circ}$ - 14.25$^{\circ}$ twisted structures. The insets show the detailed structure of a few states at the top of the valence moir\'e bands. (f) The evolution of moir\'e band widths of twisted 2L BaTiO$_3$. We also represented band widths of twisted 2L graphene,~\cite{carr2018pressure} MoS$_2$,~\cite{xian2021realization} and In$_2$Se$_3$,~\cite{tao2022designing} for comparison.
}\label{fig4}
\end{figure}

Figure~\ref{fig4}(a-e) shows the electronic spectrum of five twisted 2L BaTiO$_3$ structures corresponding to different ${\theta}_{\rm{t}}$. 
The conduction bands are parabolic in all cases, which bears similarities to the untwisted bilayer apart from the zone folding. This indicates that ${\theta}_{\rm{t}}$ does not affect the conduction states in a significant manner. 

To understand the evolution of the valence band with ${\theta}_{\rm{t}}$, we zoom into a smaller energy window in the associated insets.
Remarkably, our results indicate the presence of a single detached quasi-flat band at the valence band maximum, when 28.07$^{\circ}$$\leq{\theta}_{\rm{t}}\leq$16.26$^{\circ}$, although the lattice constants of the supercell structures are relatively small compared to other moir\'e system. 
To put in perspective, we summarized in in Fig.~\ref{fig4}(f) the bandwidth of twisted 2L BaTiO$_3$ and other moir\'e materials, such as twisted bilayer graphene,~\cite{carr2018pressure} MoS$_2$,~\cite{{xian2021realization}} and In$_2$Se$_3$.~\cite{tao2022designing}
Upon decreasing ${\theta}_{\rm{t}}$, the bandwidth of 2L BaTiO$_3$ initially decreases, reaches a minimum near ${\theta}_{\rm{t}}$=18.92$^{\circ}$, and then increases again.
As ${\theta}_{\rm{t}}$ decreases further, the gap between the flat band and the rest of the valence bands closes, giving rise to a multi-fold degeneracy at M point, as apparent in the ${\theta}_{\rm{t}}$=14.25$^{\circ}$ twisted structure.
In other words, the ``magic'' angle of 2L BaTiO$_3$, where the band flatness is maximized, is predicted to be around 18.92$^{\circ}$.
More interestingly, the bandwidth of the flat band at ${\theta}_{\rm{t}}$=18.92$^{\circ}$ is evaluated to be 5.5~meV, which is comparable with those of the magic angle 2L graphene ($\sim$10~meV at ${\theta}_{\rm{t}}$$\sim$1$^{\circ}$)~\cite{cao2018correlated,cao2018unconventional,carr2018pressure}, 2L MoS$_2$ ($\sim$11~meV at ${\theta}_{\rm{t}}$$\sim$2$^{\circ}$)~\cite{xian2021realization}, and 2L In$_2$Se$_3$ ($\sim$0.9~meV at ${\theta}_{\rm{t}}$=9.43$^{\circ}$). These results indicate that strong correlation effects might become relevant in twisted oxide membranes.

\begin{figure}[t]
\includegraphics[width=1.00\columnwidth]{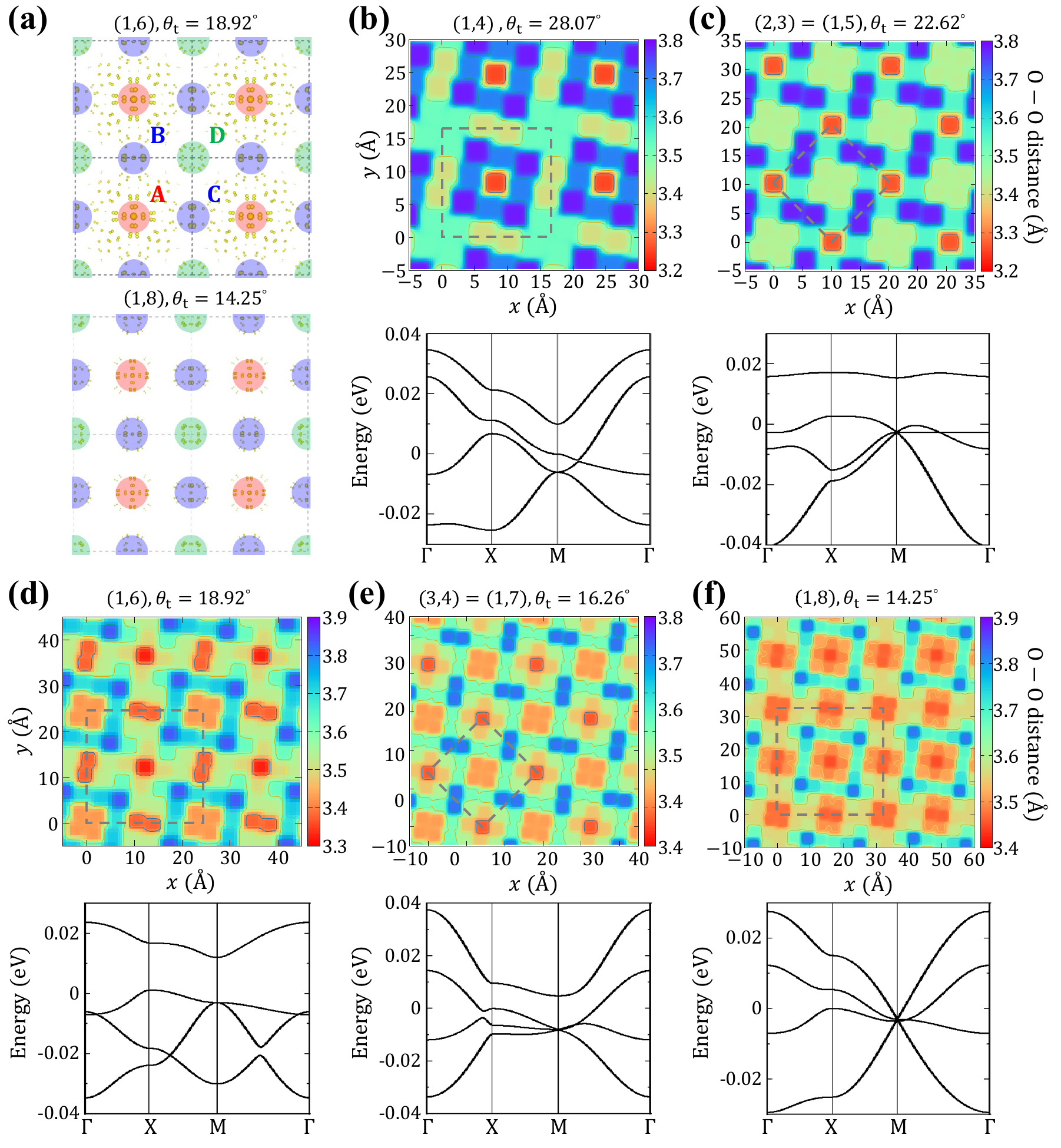}
\caption{\textbf{Topological Lieb lattice in twisted bilayer BaTiO$_3$}
Lattice model describing the dispersion of states localized at unequal vortex sites. (a) The charge density of top of valence band states and related tight-binding lattice are displayed. The tight-binding sites are labelled A, B, C and D, as shown in the upper panel. (b-f) Evolution of the tight-binding bands with twist angle. At this level of description, the twist angle dependence is assumed to mainly impact the onsite energies of the tight-binding sites. The onsite energies are qualitatively estimated from the O-O interlayer distances at the vortex sites, as shown in the colormaps. Grey dashed boxes show the primitive unitcell structure in each twist angle.}
\label{fig5}
\end{figure}

To better analyse the electron valence states in twisted 2L BaTiO$_3$, we investigated the associated charge density distribution in real space. We show the charge density of the topmost valence band for the ${\theta}_{\rm{t}}$=14.25$^{\circ}$ twisted structure in the top panel of Fig.~\ref{fig5}(a). We found the existence of localized charges at the local stacking regions, ``A'' (0,0) , ``B'' (0.5,0), ``C'' (0,0.5), and ``D'' (0.5,0.5) sites as shown in the bottom panel Fig.~\ref{fig5}(a), within the moir\'e supercell. These regions are equivalent to the four possible vortex configurations shown, for instance, in Fig.~\ref{fig2}(g) and associated with distinct potential strengths shown in Fig.~\ref{fig3}(e), indicating the intimate connection between the moir\'e vortex pattern and localized electronic states.

The above considerations motivate us to construct a tight-binding model for the localized states, in a lattice set by these four ``atomic sites", to provide qualitative insights in regards to some of the features of the topmost valence bands shown in Fig.~\ref{fig4}. The model Hamiltonian describing such electron states becomes
\begin{eqnarray}
H(\textbf{k}) = \left[
\begin{tabular}{cccc}
     $\epsilon_A$ & $t_{AB}f_y(\textbf{k})$ & $t_{AC}f_x(\textbf{k})$ & $t_{AD}g(\textbf{k})$ \\
     $t_{AB}f_y(\textbf{k})$ & $\epsilon_B$ & $t_{BC}g(\textbf{k})$ & $t_{BD}f_x(\textbf{k})$\\
     $t_{AC}f_x(\textbf{k})$ & $t_{BC}g(\textbf{k})$ & $\epsilon_C$ & $t_{CD}f_y(\textbf{k})$\\
     $t_{AD}g(\textbf{k})$ & $t_{BD}f_x(\textbf{k})$ & $t_{CD}f_y(\textbf{k})$ & $\epsilon_D$ 
\end{tabular}
\right], \nonumber \\
    \label{Hamiltonian}
\end{eqnarray}
where $\epsilon_{j}(\textbf{k}) = \epsilon_j - 2t^j_{NNN} \cos(k_x a) - 2t^j_{NNN} \cos(k_y a)$, with $j = A,B,C,D$ and $\epsilon_j$ refers to the onsite energies while $t_{ij}$ ($t^{j}_{NNN}$) describes 
 nearest-neighbor (next nearest-neighbor) hopping processes between neighboring sites. The structure factors corresponding to the lattice configuration are, $f_{x,y}(\textbf{k}) = 2\cos(k_{x,y} a/2)$ and $g(\textbf{k}) = 2\cos[(k_x + k_y) a/2] + 2\cos[(k_x - k_y) a/2]$, where $a$ is the lattice constant of the moir\'e unit cell. 
Because sites ``B'' and ``C'' are related by a $C_{4z}$ operation, we have $\epsilon_B = \epsilon_C$, $t_{AB} = t_{AC}$ ($t_{BD} = t_{CD}$) throughout. All model parameters are given in Note~\ref{note1} in SI.

The surface potential map shown in Figure~\ref{fig3}(e) gives a hint on how to constrain the onsite energies of our tight-binding model. In that case, the $18.92^{\circ}$ twisted structure is characterized by O regions with different onsite energies, which roughly obey $|\epsilon_A| > |\epsilon_{BC}| >\epsilon_{D}$. We note that such spatially dependent potential profile is connected to the O-O interlayer distances within the moir\'e unit cell, which is twist angle dependent. We found that regions with larger potentials are associated with smaller O-O interlayer distances. 
Therefore, we quantitatively interpreted the O-O interlayer distances as being related to the onsite energies in our tight-binding approach, which we address next.

The tight-binding electronic structures, corresponding to onsite energy configurations qualitatively similar to surface potential profiles, are shown in Fig.~\ref{fig5}(b-f), along with the interlayer O-O distance maps. We observe that distinct onsite energies induced by varying local stacking orders in the twisted BaTiO$_3$ bilayers will naturally cause the upper band, describing localized states at the ``A'' site, to separate from the lower bands while reducing its band width, for all cases except at $\theta_t < 14.25^{\circ}$ twisted lattices. This is consistent with the features observed in our first principles calculations, where a detached quasi-flat band appears in the spectrum whenever the local $A$ site energy exceeds $\epsilon_B$, $\epsilon_C$ and $\epsilon_D$, as apparent from the minimum O-O interlayer distances.

Note that the remaining set of tight-binding bands are in good qualitative agreement with our first principles calculations as well, e.g., band crossing at $M$ and dispersion throughout the moir\'e Brillouin zone. As the twist angle is decreased further below $16.92^{\circ}$, the onsite energy $\epsilon_A$ approaches $\epsilon_{B,C,D}$ from above. 
Hence, we assume next an onsite energy configuration qualitatively similar to that of the $14.25^{\circ}$ structure, i.e. $|\epsilon_A| \approx |\epsilon_{BC}| > \epsilon_{D}$. 
We note that a multi-fold degeneracy at the $M$-point emerges naturally in this configuration with two linearly crossing bands, in good qualitative agreement with our first principles results shown in Fig.~\ref{fig4}(f).
Such feature on a 2D square lattice is a hallmark of electron states in a topological Lieb lattice.~\cite{jiang2019topological,jiang2020topological,cui2020realization} 
In fact, Hamiltonian~(\ref{Hamiltonian}) describes two interpenetrating Lieb lattices sharing ``edge" sites ($B, C$), but with distinct central sites ($A$ or $D$). 
An ``ideal" Lieb lattice can be obtained from Hamiltonian~(\ref{Hamiltonian}) by taking, for instance, $|\epsilon_A| \gg |\epsilon_{B,C,D}|$, in which case the $A$ site can be ``factored" out through L\"owdin partitioning, leaving behind a three-site Lieb lattice configuration. Hence, electron states in twisted BaTiO$_3$ bilayers configure distorted interpenetrating Lieb lattice with twist angle-dependent hopping and onsite energies, a crucial step towards the realization of alternative topological electronic states in condensed matter.

\section{Conclusion}

In summary, using first-principles calculations based on density functional theory, we investigated crystal and electronic structures of twisted BaTiO$_3$ bilayer.
Among three possible interface configurations in twisted 2L BaTiO$_3$, 
we found that BaO-BaO interface resembles the vdW interface but exhibits a one to two orders of magnitude larger GSFE compared to other mori\'{e} systems.
While the conduction bands are insensitive to interlayer interactions, the valence bands are greatly influenced by the interlayer hopping between oxygen $p$ orbitals, highlighting the importance of interlayer coupling.

With a non-zero ${\theta}_{\rm{t}}$, the huge GSFE induces substantial relaxation within a layer, resulting in chiral in-plane vortex patterns that were recently observed experimentally, with opposite chirality in the two layers.
The interlayer interactions also gave rise to local out-of-plane dipole moments and robust surface electrostatic potentials. 
More intriguingly, we found an isolated flat band at the valence band maximum over a wide range of ${\theta}_{\rm{t}}$.
The bandwidth of the flat bands near ${\theta}_{\rm{t}}$$\sim$19$^{\circ}$ is comparable with those of magic-angle bilayer graphene, indicating strong correlation effects in our system. 

We further constructed a tight-binding lattice model based on the localized states associated with the topmost valence bands. The localized state sites correspond to the four inequivalent local in-plane vortex sites. Our model provides qualitative insights as to the origins of the detached quasi-flat bands observed in our first principles calculations. 
These bands are well described by two interpenetrating Lieb lattice, with twist angle-dependent onsite energies. These observations indicate that twisted 2L BaTiO$_3$ host unique electronic behavior, offering a new perspective towards what might be the first realization of an inorganic and electronic Lieb lattice. 

\section{method}
The first-principles DFT calculation~\cite{Kohn1965} was carried out using Vienna \textit{ab initio} simulation package (VASP).~\cite{Kresse1996} We employed the plane wave basis to expand the electronic wavefunctions with a kinetic energy cutoff of 500~eV. 
The projector-augmented wave pseudopotentials~\cite{{Blochl1994},{Kresse1999}} were used for the valence electrons, and the exchange-correlation (XC) functional was treated within the generalized gradient approximation of Perdew–Burke–Ernzerhof (PBE).~\cite{Perdew1996}
The van der Waals interaction between two adjacent layer was treated by Grimme D3 methods~\cite{grimme-d3}.
A sufficiently large vacuum region ($>15~\textrm{\AA}$) was included in the unitcell to avoid any spurious interlayer interactions.
The atomic basis was carefully relaxed until the Helmann-Feynman force acting on every atom was smaller than 0.01~eV/$\textrm{\AA}$.
The 15$\times$15$\times$1 $k$-mesh was used to sample the Brillouin zone for untwisted structures and $\Gamma$ point sampling was used for all other twisted structures.
The equilibrium lattice constant of monolayer BaTiO$_3$ was calculated to be 4.002~{\AA}, and the same value was used for all other structures.

\textit{Acknowledgments} 
S. L. is supported by Basic Science Research Program through the National Research Foundation of Korea funded by the Ministry of Education (NRF-2021R1A6A3A14038837).
S. L. and T. L. are partially supported by NSF DMREF-1921629. D. S and T. L. acknowledge partial support  from Office of Naval Research MURI grant N00014-23-1-2567. B. J. acknowledges support by the Air Force Office of Scientific Research (AFOSR) through Grant Nos. FA9550-21-1-0025 and FA9550-21-0460.
We thank Richard D. James for helpful comments.


\begin{thebibliography}{43}%
\makeatletter
\providecommand \@ifxundefined [1]{%
 \@ifx{#1\undefined}
}%
\providecommand \@ifnum [1]{%
 \ifnum #1\expandafter \@firstoftwo
 \else \expandafter \@secondoftwo
 \fi
}%
\providecommand \@ifx [1]{%
 \ifx #1\expandafter \@firstoftwo
 \else \expandafter \@secondoftwo
 \fi
}%
\providecommand \natexlab [1]{#1}%
\providecommand \enquote  [1]{``#1''}%
\providecommand \bibnamefont  [1]{#1}%
\providecommand \bibfnamefont [1]{#1}%
\providecommand \citenamefont [1]{#1}%
\providecommand \href@noop [0]{\@secondoftwo}%
\providecommand \href [0]{\begingroup \@sanitize@url \@href}%
\providecommand \@href[1]{\@@startlink{#1}\@@href}%
\providecommand \@@href[1]{\endgroup#1\@@endlink}%
\providecommand \@sanitize@url [0]{\catcode `\\12\catcode `\$12\catcode `\&12\catcode `\#12\catcode `\^12\catcode `\_12\catcode `\%12\relax}%
\providecommand \@@startlink[1]{}%
\providecommand \@@endlink[0]{}%
\providecommand \url  [0]{\begingroup\@sanitize@url \@url }%
\providecommand \@url [1]{\endgroup\@href {#1}{\urlprefix }}%
\providecommand \urlprefix  [0]{URL }%
\providecommand \Eprint [0]{\href }%
\providecommand \doibase [0]{https://doi.org/}%
\providecommand \selectlanguage [0]{\@gobble}%
\providecommand \bibinfo  [0]{\@secondoftwo}%
\providecommand \bibfield  [0]{\@secondoftwo}%
\providecommand \translation [1]{[#1]}%
\providecommand \BibitemOpen [0]{}%
\providecommand \bibitemStop [0]{}%
\providecommand \bibitemNoStop [0]{.\EOS\space}%
\providecommand \EOS [0]{\spacefactor3000\relax}%
\providecommand \BibitemShut  [1]{\csname bibitem#1\endcsname}%
\let\auto@bib@innerbib\@empty
\bibitem [{\citenamefont {Carr}\ \emph {et~al.}(2020)\citenamefont {Carr}, \citenamefont {Fang},\ and\ \citenamefont {Kaxiras}}]{carr2020electronic}%
  \BibitemOpen
  \bibfield  {author} {\bibinfo {author} {\bibfnamefont {S.}~\bibnamefont {Carr}}, \bibinfo {author} {\bibfnamefont {S.}~\bibnamefont {Fang}},\ and\ \bibinfo {author} {\bibfnamefont {E.}~\bibnamefont {Kaxiras}},\ }\bibfield  {title} {\bibinfo {title} {Electronic-structure methods for twisted moir{\'e} layers},\ }\href@noop {} {\bibfield  {journal} {\bibinfo  {journal} {Nat. Rev. Mater.}\ }\textbf {\bibinfo {volume} {5}},\ \bibinfo {pages} {748} (\bibinfo {year} {2020})}\BibitemShut {NoStop}%
\bibitem [{\citenamefont {Andrei}\ and\ \citenamefont {MacDonald}(2020)}]{andrei2020graphene}%
  \BibitemOpen
  \bibfield  {author} {\bibinfo {author} {\bibfnamefont {E.~Y.}\ \bibnamefont {Andrei}}\ and\ \bibinfo {author} {\bibfnamefont {A.~H.}\ \bibnamefont {MacDonald}},\ }\bibfield  {title} {\bibinfo {title} {Graphene bilayers with a twist},\ }\href@noop {} {\bibfield  {journal} {\bibinfo  {journal} {Nat. Mater.}\ }\textbf {\bibinfo {volume} {19}},\ \bibinfo {pages} {1265} (\bibinfo {year} {2020})}\BibitemShut {NoStop}%
\bibitem [{\citenamefont {Bistritzer}\ and\ \citenamefont {MacDonald}(2011)}]{bistritzer2011moire}%
  \BibitemOpen
  \bibfield  {author} {\bibinfo {author} {\bibfnamefont {R.}~\bibnamefont {Bistritzer}}\ and\ \bibinfo {author} {\bibfnamefont {A.~H.}\ \bibnamefont {MacDonald}},\ }\bibfield  {title} {\bibinfo {title} {Moir{\'e} bands in twisted double-layer graphene},\ }\href@noop {} {\bibfield  {journal} {\bibinfo  {journal} {Proc. Natl. Acad. Sci.}\ }\textbf {\bibinfo {volume} {108}},\ \bibinfo {pages} {12233} (\bibinfo {year} {2011})}\BibitemShut {NoStop}%
\bibitem [{\citenamefont {Cao}\ \emph {et~al.}(2018{\natexlab{a}})\citenamefont {Cao}, \citenamefont {Fatemi}, \citenamefont {Demir}, \citenamefont {Fang}, \citenamefont {Tomarken}, \citenamefont {Luo}, \citenamefont {Sanchez-Yamagishi}, \citenamefont {Watanabe}, \citenamefont {Taniguchi}, \citenamefont {Kaxiras} \emph {et~al.}}]{cao2018correlated}%
  \BibitemOpen
  \bibfield  {author} {\bibinfo {author} {\bibfnamefont {Y.}~\bibnamefont {Cao}}, \bibinfo {author} {\bibfnamefont {V.}~\bibnamefont {Fatemi}}, \bibinfo {author} {\bibfnamefont {A.}~\bibnamefont {Demir}}, \bibinfo {author} {\bibfnamefont {S.}~\bibnamefont {Fang}}, \bibinfo {author} {\bibfnamefont {S.~L.}\ \bibnamefont {Tomarken}}, \bibinfo {author} {\bibfnamefont {J.~Y.}\ \bibnamefont {Luo}}, \bibinfo {author} {\bibfnamefont {J.~D.}\ \bibnamefont {Sanchez-Yamagishi}}, \bibinfo {author} {\bibfnamefont {K.}~\bibnamefont {Watanabe}}, \bibinfo {author} {\bibfnamefont {T.}~\bibnamefont {Taniguchi}}, \bibinfo {author} {\bibfnamefont {E.}~\bibnamefont {Kaxiras}}, \emph {et~al.},\ }\bibfield  {title} {\bibinfo {title} {Correlated insulator behaviour at half-filling in magic-angle graphene superlattices},\ }\href@noop {} {\bibfield  {journal} {\bibinfo  {journal} {Nature}\ }\textbf {\bibinfo {volume} {556}},\ \bibinfo {pages} {80} (\bibinfo {year} {2018}{\natexlab{a}})}\BibitemShut {NoStop}%
\bibitem [{\citenamefont {Cao}\ \emph {et~al.}(2018{\natexlab{b}})\citenamefont {Cao}, \citenamefont {Fatemi}, \citenamefont {Fang}, \citenamefont {Watanabe}, \citenamefont {Taniguchi}, \citenamefont {Kaxiras},\ and\ \citenamefont {Jarillo-Herrero}}]{cao2018unconventional}%
  \BibitemOpen
  \bibfield  {author} {\bibinfo {author} {\bibfnamefont {Y.}~\bibnamefont {Cao}}, \bibinfo {author} {\bibfnamefont {V.}~\bibnamefont {Fatemi}}, \bibinfo {author} {\bibfnamefont {S.}~\bibnamefont {Fang}}, \bibinfo {author} {\bibfnamefont {K.}~\bibnamefont {Watanabe}}, \bibinfo {author} {\bibfnamefont {T.}~\bibnamefont {Taniguchi}}, \bibinfo {author} {\bibfnamefont {E.}~\bibnamefont {Kaxiras}},\ and\ \bibinfo {author} {\bibfnamefont {P.}~\bibnamefont {Jarillo-Herrero}},\ }\bibfield  {title} {\bibinfo {title} {Unconventional superconductivity in magic-angle graphene superlattices},\ }\href@noop {} {\bibfield  {journal} {\bibinfo  {journal} {Nature}\ }\textbf {\bibinfo {volume} {556}},\ \bibinfo {pages} {43} (\bibinfo {year} {2018}{\natexlab{b}})}\BibitemShut {NoStop}%
\bibitem [{\citenamefont {Wu}\ \emph {et~al.}(2018)\citenamefont {Wu}, \citenamefont {Lovorn}, \citenamefont {Tutuc},\ and\ \citenamefont {MacDonald}}]{wu2018hubbard}%
  \BibitemOpen
  \bibfield  {author} {\bibinfo {author} {\bibfnamefont {F.}~\bibnamefont {Wu}}, \bibinfo {author} {\bibfnamefont {T.}~\bibnamefont {Lovorn}}, \bibinfo {author} {\bibfnamefont {E.}~\bibnamefont {Tutuc}},\ and\ \bibinfo {author} {\bibfnamefont {A.~H.}\ \bibnamefont {MacDonald}},\ }\bibfield  {title} {\bibinfo {title} {Hubbard model physics in transition metal dichalcogenide moir{\'e} bands},\ }\href@noop {} {\bibfield  {journal} {\bibinfo  {journal} {Phys. Rev. Lett.}\ }\textbf {\bibinfo {volume} {121}},\ \bibinfo {pages} {026402} (\bibinfo {year} {2018})}\BibitemShut {NoStop}%
\bibitem [{\citenamefont {Naik}\ and\ \citenamefont {Jain}(2018)}]{naik2018ultraflatbands}%
  \BibitemOpen
  \bibfield  {author} {\bibinfo {author} {\bibfnamefont {M.~H.}\ \bibnamefont {Naik}}\ and\ \bibinfo {author} {\bibfnamefont {M.}~\bibnamefont {Jain}},\ }\bibfield  {title} {\bibinfo {title} {Ultraflatbands and shear solitons in moir{\'e} patterns of twisted bilayer transition metal dichalcogenides},\ }\href@noop {} {\bibfield  {journal} {\bibinfo  {journal} {Phys. Rev. Lett.}\ }\textbf {\bibinfo {volume} {121}},\ \bibinfo {pages} {266401} (\bibinfo {year} {2018})}\BibitemShut {NoStop}%
\bibitem [{\citenamefont {Xian}\ \emph {et~al.}(2019)\citenamefont {Xian}, \citenamefont {Kennes}, \citenamefont {Tancogne-Dejean}, \citenamefont {Altarelli},\ and\ \citenamefont {Rubio}}]{xian2019multiflat}%
  \BibitemOpen
  \bibfield  {author} {\bibinfo {author} {\bibfnamefont {L.}~\bibnamefont {Xian}}, \bibinfo {author} {\bibfnamefont {D.~M.}\ \bibnamefont {Kennes}}, \bibinfo {author} {\bibfnamefont {N.}~\bibnamefont {Tancogne-Dejean}}, \bibinfo {author} {\bibfnamefont {M.}~\bibnamefont {Altarelli}},\ and\ \bibinfo {author} {\bibfnamefont {A.}~\bibnamefont {Rubio}},\ }\bibfield  {title} {\bibinfo {title} {Multiflat bands and strong correlations in twisted bilayer boron nitride: Doping-induced correlated insulator and superconductor},\ }\href@noop {} {\bibfield  {journal} {\bibinfo  {journal} {Nano Lett.}\ }\textbf {\bibinfo {volume} {19}},\ \bibinfo {pages} {4934} (\bibinfo {year} {2019})}\BibitemShut {NoStop}%
\bibitem [{\citenamefont {Kim}\ \emph {et~al.}(2017)\citenamefont {Kim}, \citenamefont {Cruz}, \citenamefont {Lee}, \citenamefont {Alawode}, \citenamefont {Choi}, \citenamefont {Song}, \citenamefont {Johnson}, \citenamefont {Heidelberger}, \citenamefont {Kong}, \citenamefont {Choi} \emph {et~al.}}]{kim2017remote}%
  \BibitemOpen
  \bibfield  {author} {\bibinfo {author} {\bibfnamefont {Y.}~\bibnamefont {Kim}}, \bibinfo {author} {\bibfnamefont {S.~S.}\ \bibnamefont {Cruz}}, \bibinfo {author} {\bibfnamefont {K.}~\bibnamefont {Lee}}, \bibinfo {author} {\bibfnamefont {B.~O.}\ \bibnamefont {Alawode}}, \bibinfo {author} {\bibfnamefont {C.}~\bibnamefont {Choi}}, \bibinfo {author} {\bibfnamefont {Y.}~\bibnamefont {Song}}, \bibinfo {author} {\bibfnamefont {J.~M.}\ \bibnamefont {Johnson}}, \bibinfo {author} {\bibfnamefont {C.}~\bibnamefont {Heidelberger}}, \bibinfo {author} {\bibfnamefont {W.}~\bibnamefont {Kong}}, \bibinfo {author} {\bibfnamefont {S.}~\bibnamefont {Choi}}, \emph {et~al.},\ }\bibfield  {title} {\bibinfo {title} {Remote epitaxy through graphene enables two-dimensional material-based layer transfer},\ }\href@noop {} {\bibfield  {journal} {\bibinfo  {journal} {Nature}\ }\textbf {\bibinfo {volume} {544}},\ \bibinfo {pages} {340} (\bibinfo {year} {2017})}\BibitemShut {NoStop}%
\bibitem [{\citenamefont {Kim}\ \emph {et~al.}(2022{\natexlab{a}})\citenamefont {Kim}, \citenamefont {Chang}, \citenamefont {Lee}, \citenamefont {Jiang}, \citenamefont {Jeong}, \citenamefont {Park}, \citenamefont {Meng}, \citenamefont {Ji}, \citenamefont {Kwon}, \citenamefont {Sun} \emph {et~al.}}]{kim2022remote}%
  \BibitemOpen
  \bibfield  {author} {\bibinfo {author} {\bibfnamefont {H.}~\bibnamefont {Kim}}, \bibinfo {author} {\bibfnamefont {C.~S.}\ \bibnamefont {Chang}}, \bibinfo {author} {\bibfnamefont {S.}~\bibnamefont {Lee}}, \bibinfo {author} {\bibfnamefont {J.}~\bibnamefont {Jiang}}, \bibinfo {author} {\bibfnamefont {J.}~\bibnamefont {Jeong}}, \bibinfo {author} {\bibfnamefont {M.}~\bibnamefont {Park}}, \bibinfo {author} {\bibfnamefont {Y.}~\bibnamefont {Meng}}, \bibinfo {author} {\bibfnamefont {J.}~\bibnamefont {Ji}}, \bibinfo {author} {\bibfnamefont {Y.}~\bibnamefont {Kwon}}, \bibinfo {author} {\bibfnamefont {X.}~\bibnamefont {Sun}}, \emph {et~al.},\ }\bibfield  {title} {\bibinfo {title} {Remote epitaxy},\ }\href@noop {} {\bibfield  {journal} {\bibinfo  {journal} {Nat. Rev. Methods Primers}\ }\textbf {\bibinfo {volume} {2}},\ \bibinfo {pages} {40} (\bibinfo {year} {2022}{\natexlab{a}})}\BibitemShut {NoStop}%
\bibitem [{\citenamefont {Manzo}\ \emph {et~al.}(2022)\citenamefont {Manzo}, \citenamefont {Strohbeen}, \citenamefont {Lim}, \citenamefont {Saraswat}, \citenamefont {Du}, \citenamefont {Xu}, \citenamefont {Pokharel}, \citenamefont {Mawst}, \citenamefont {Arnold},\ and\ \citenamefont {Kawasaki}}]{manzo2022pinhole}%
  \BibitemOpen
  \bibfield  {author} {\bibinfo {author} {\bibfnamefont {S.}~\bibnamefont {Manzo}}, \bibinfo {author} {\bibfnamefont {P.~J.}\ \bibnamefont {Strohbeen}}, \bibinfo {author} {\bibfnamefont {Z.~H.}\ \bibnamefont {Lim}}, \bibinfo {author} {\bibfnamefont {V.}~\bibnamefont {Saraswat}}, \bibinfo {author} {\bibfnamefont {D.}~\bibnamefont {Du}}, \bibinfo {author} {\bibfnamefont {S.}~\bibnamefont {Xu}}, \bibinfo {author} {\bibfnamefont {N.}~\bibnamefont {Pokharel}}, \bibinfo {author} {\bibfnamefont {L.~J.}\ \bibnamefont {Mawst}}, \bibinfo {author} {\bibfnamefont {M.~S.}\ \bibnamefont {Arnold}},\ and\ \bibinfo {author} {\bibfnamefont {J.~K.}\ \bibnamefont {Kawasaki}},\ }\bibfield  {title} {\bibinfo {title} {{Pinhole-seeded lateral epitaxy and exfoliation of GaSb films on graphene-terminated surfaces}},\ }\href@noop {} {\bibfield  {journal} {\bibinfo  {journal} {Nat. Commun.}\ }\textbf {\bibinfo {volume} {13}},\ \bibinfo {pages} {4014} (\bibinfo {year} {2022})}\BibitemShut {NoStop}%
\bibitem [{\citenamefont {Jang}\ \emph {et~al.}(2023)\citenamefont {Jang}, \citenamefont {Ahn}, \citenamefont {Lee}, \citenamefont {Lee}, \citenamefont {Lee}, \citenamefont {Kim}, \citenamefont {Kim}, \citenamefont {Park}, \citenamefont {Kwon}, \citenamefont {Choi} \emph {et~al.}}]{jang2023thru}%
  \BibitemOpen
  \bibfield  {author} {\bibinfo {author} {\bibfnamefont {D.}~\bibnamefont {Jang}}, \bibinfo {author} {\bibfnamefont {C.}~\bibnamefont {Ahn}}, \bibinfo {author} {\bibfnamefont {Y.}~\bibnamefont {Lee}}, \bibinfo {author} {\bibfnamefont {S.}~\bibnamefont {Lee}}, \bibinfo {author} {\bibfnamefont {H.}~\bibnamefont {Lee}}, \bibinfo {author} {\bibfnamefont {D.}~\bibnamefont {Kim}}, \bibinfo {author} {\bibfnamefont {Y.}~\bibnamefont {Kim}}, \bibinfo {author} {\bibfnamefont {J.-Y.}\ \bibnamefont {Park}}, \bibinfo {author} {\bibfnamefont {Y.-K.}\ \bibnamefont {Kwon}}, \bibinfo {author} {\bibfnamefont {J.}~\bibnamefont {Choi}}, \emph {et~al.},\ }\bibfield  {title} {\bibinfo {title} {Thru-hole epitaxy: A highway for controllable and transferable epitaxial growth},\ }\href@noop {} {\bibfield  {journal} {\bibinfo  {journal} {Adv. Mater. Interfaces}\ }\textbf {\bibinfo {volume} {10}},\ \bibinfo {pages} {2201406} (\bibinfo {year} {2023})}\BibitemShut {NoStop}%
\bibitem [{\citenamefont {Yoon}\ \emph {et~al.}(2022)\citenamefont {Yoon}, \citenamefont {Truttmann}, \citenamefont {Liu}, \citenamefont {Matthews}, \citenamefont {Choo}, \citenamefont {Su}, \citenamefont {Saraswat}, \citenamefont {Manzo}, \citenamefont {Arnold}, \citenamefont {Bowden} \emph {et~al.}}]{yoon2022freestanding}%
  \BibitemOpen
  \bibfield  {author} {\bibinfo {author} {\bibfnamefont {H.}~\bibnamefont {Yoon}}, \bibinfo {author} {\bibfnamefont {T.~K.}\ \bibnamefont {Truttmann}}, \bibinfo {author} {\bibfnamefont {F.}~\bibnamefont {Liu}}, \bibinfo {author} {\bibfnamefont {B.~E.}\ \bibnamefont {Matthews}}, \bibinfo {author} {\bibfnamefont {S.}~\bibnamefont {Choo}}, \bibinfo {author} {\bibfnamefont {Q.}~\bibnamefont {Su}}, \bibinfo {author} {\bibfnamefont {V.}~\bibnamefont {Saraswat}}, \bibinfo {author} {\bibfnamefont {S.}~\bibnamefont {Manzo}}, \bibinfo {author} {\bibfnamefont {M.~S.}\ \bibnamefont {Arnold}}, \bibinfo {author} {\bibfnamefont {M.~E.}\ \bibnamefont {Bowden}}, \emph {et~al.},\ }\bibfield  {title} {\bibinfo {title} {Freestanding epitaxial srtio3 nanomembranes via remote epitaxy using hybrid molecular beam epitaxy},\ }\href@noop {} {\bibfield  {journal} {\bibinfo  {journal} {Sci. Adv.}\ }\textbf {\bibinfo {volume} {8}},\ \bibinfo {pages} {eadd5328} (\bibinfo {year} {2022})}\BibitemShut {NoStop}%
\bibitem [{\citenamefont {Lu}\ \emph {et~al.}(2016)\citenamefont {Lu}, \citenamefont {Baek}, \citenamefont {Hong}, \citenamefont {Kourkoutis}, \citenamefont {Hikita},\ and\ \citenamefont {Hwang}}]{lu2016synthesis}%
  \BibitemOpen
  \bibfield  {author} {\bibinfo {author} {\bibfnamefont {D.}~\bibnamefont {Lu}}, \bibinfo {author} {\bibfnamefont {D.~J.}\ \bibnamefont {Baek}}, \bibinfo {author} {\bibfnamefont {S.~S.}\ \bibnamefont {Hong}}, \bibinfo {author} {\bibfnamefont {L.~F.}\ \bibnamefont {Kourkoutis}}, \bibinfo {author} {\bibfnamefont {Y.}~\bibnamefont {Hikita}},\ and\ \bibinfo {author} {\bibfnamefont {H.~Y.}\ \bibnamefont {Hwang}},\ }\bibfield  {title} {\bibinfo {title} {Synthesis of freestanding single-crystal perovskite films and heterostructures by etching of sacrificial water-soluble layers},\ }\href@noop {} {\bibfield  {journal} {\bibinfo  {journal} {Nat. Mater.}\ }\textbf {\bibinfo {volume} {15}},\ \bibinfo {pages} {1255} (\bibinfo {year} {2016})}\BibitemShut {NoStop}%
\bibitem [{\citenamefont {Hong}\ \emph {et~al.}(2017)\citenamefont {Hong}, \citenamefont {Yu}, \citenamefont {Lu}, \citenamefont {Marshall}, \citenamefont {Hikita}, \citenamefont {Cui},\ and\ \citenamefont {Hwang}}]{hong2017two}%
  \BibitemOpen
  \bibfield  {author} {\bibinfo {author} {\bibfnamefont {S.~S.}\ \bibnamefont {Hong}}, \bibinfo {author} {\bibfnamefont {J.~H.}\ \bibnamefont {Yu}}, \bibinfo {author} {\bibfnamefont {D.}~\bibnamefont {Lu}}, \bibinfo {author} {\bibfnamefont {A.~F.}\ \bibnamefont {Marshall}}, \bibinfo {author} {\bibfnamefont {Y.}~\bibnamefont {Hikita}}, \bibinfo {author} {\bibfnamefont {Y.}~\bibnamefont {Cui}},\ and\ \bibinfo {author} {\bibfnamefont {H.~Y.}\ \bibnamefont {Hwang}},\ }\bibfield  {title} {\bibinfo {title} {Two-dimensional limit of crystalline order in perovskite membrane films},\ }\href@noop {} {\bibfield  {journal} {\bibinfo  {journal} {Sci. Adv.}\ }\textbf {\bibinfo {volume} {3}},\ \bibinfo {pages} {eaao5173} (\bibinfo {year} {2017})}\BibitemShut {NoStop}%
\bibitem [{\citenamefont {Ji}\ \emph {et~al.}(2019)\citenamefont {Ji}, \citenamefont {Cai}, \citenamefont {Paudel}, \citenamefont {Sun}, \citenamefont {Zhang}, \citenamefont {Han}, \citenamefont {Wei}, \citenamefont {Zang}, \citenamefont {Gu}, \citenamefont {Zhang} \emph {et~al.}}]{ji2019freestanding}%
  \BibitemOpen
  \bibfield  {author} {\bibinfo {author} {\bibfnamefont {D.}~\bibnamefont {Ji}}, \bibinfo {author} {\bibfnamefont {S.}~\bibnamefont {Cai}}, \bibinfo {author} {\bibfnamefont {T.~R.}\ \bibnamefont {Paudel}}, \bibinfo {author} {\bibfnamefont {H.}~\bibnamefont {Sun}}, \bibinfo {author} {\bibfnamefont {C.}~\bibnamefont {Zhang}}, \bibinfo {author} {\bibfnamefont {L.}~\bibnamefont {Han}}, \bibinfo {author} {\bibfnamefont {Y.}~\bibnamefont {Wei}}, \bibinfo {author} {\bibfnamefont {Y.}~\bibnamefont {Zang}}, \bibinfo {author} {\bibfnamefont {M.}~\bibnamefont {Gu}}, \bibinfo {author} {\bibfnamefont {Y.}~\bibnamefont {Zhang}}, \emph {et~al.},\ }\bibfield  {title} {\bibinfo {title} {Freestanding crystalline oxide perovskites down to the monolayer limit},\ }\href@noop {} {\bibfield  {journal} {\bibinfo  {journal} {Nature}\ }\textbf {\bibinfo {volume} {570}},\ \bibinfo {pages} {87} (\bibinfo {year} {2019})}\BibitemShut {NoStop}%
\bibitem [{\citenamefont {Varshney}\ \emph {et~al.}(2024)\citenamefont {Varshney}, \citenamefont {Choo}, \citenamefont {Thompson}, \citenamefont {Yang}, \citenamefont {Shah}, \citenamefont {Wen}, \citenamefont {Koester}, \citenamefont {Mkhoyan}, \citenamefont {McLeod},\ and\ \citenamefont {Jalan}}]{varshney2024hybrid}%
  \BibitemOpen
  \bibfield  {author} {\bibinfo {author} {\bibfnamefont {S.}~\bibnamefont {Varshney}}, \bibinfo {author} {\bibfnamefont {S.}~\bibnamefont {Choo}}, \bibinfo {author} {\bibfnamefont {L.}~\bibnamefont {Thompson}}, \bibinfo {author} {\bibfnamefont {Z.}~\bibnamefont {Yang}}, \bibinfo {author} {\bibfnamefont {J.}~\bibnamefont {Shah}}, \bibinfo {author} {\bibfnamefont {J.}~\bibnamefont {Wen}}, \bibinfo {author} {\bibfnamefont {S.~J.}\ \bibnamefont {Koester}}, \bibinfo {author} {\bibfnamefont {K.~A.}\ \bibnamefont {Mkhoyan}}, \bibinfo {author} {\bibfnamefont {A.~S.}\ \bibnamefont {McLeod}},\ and\ \bibinfo {author} {\bibfnamefont {B.}~\bibnamefont {Jalan}},\ }\bibfield  {title} {\bibinfo {title} {Hybrid molecular beam epitaxy for single-crystalline oxide membranes with binary oxide sacrificial layers},\ }\href@noop {} {\bibfield  {journal} {\bibinfo  {journal} {ACS Nano}\ } (\bibinfo {year} {2024})}\BibitemShut {NoStop}%
\bibitem [{\citenamefont {Shen}\ \emph {et~al.}(2022)\citenamefont {Shen}, \citenamefont {Dong}, \citenamefont {Qi}, \citenamefont {Zhang}, \citenamefont {Zhu}, \citenamefont {Wu},\ and\ \citenamefont {Li}}]{shen2022observation}%
  \BibitemOpen
  \bibfield  {author} {\bibinfo {author} {\bibfnamefont {J.}~\bibnamefont {Shen}}, \bibinfo {author} {\bibfnamefont {Z.}~\bibnamefont {Dong}}, \bibinfo {author} {\bibfnamefont {M.}~\bibnamefont {Qi}}, \bibinfo {author} {\bibfnamefont {Y.}~\bibnamefont {Zhang}}, \bibinfo {author} {\bibfnamefont {C.}~\bibnamefont {Zhu}}, \bibinfo {author} {\bibfnamefont {Z.}~\bibnamefont {Wu}},\ and\ \bibinfo {author} {\bibfnamefont {D.}~\bibnamefont {Li}},\ }\bibfield  {title} {\bibinfo {title} {Observation of moir{\'e} patterns in twisted stacks of bilayer perovskite oxide nanomembranes with various lattice symmetries},\ }\href@noop {} {\bibfield  {journal} {\bibinfo  {journal} {ACS Appl. Mater. Interfaces.}\ }\textbf {\bibinfo {volume} {14}},\ \bibinfo {pages} {50386} (\bibinfo {year} {2022})}\BibitemShut {NoStop}%
\bibitem [{\citenamefont {Li}\ \emph {et~al.}(2022)\citenamefont {Li}, \citenamefont {Xiang}, \citenamefont {Chiabrera}, \citenamefont {Yun}, \citenamefont {Zhang}, \citenamefont {Kelly}, \citenamefont {Dahm}, \citenamefont {Kirchert}, \citenamefont {Cozannet}, \citenamefont {Trier} \emph {et~al.}}]{li2022stacking}%
  \BibitemOpen
  \bibfield  {author} {\bibinfo {author} {\bibfnamefont {Y.}~\bibnamefont {Li}}, \bibinfo {author} {\bibfnamefont {C.}~\bibnamefont {Xiang}}, \bibinfo {author} {\bibfnamefont {F.~M.}\ \bibnamefont {Chiabrera}}, \bibinfo {author} {\bibfnamefont {S.}~\bibnamefont {Yun}}, \bibinfo {author} {\bibfnamefont {H.}~\bibnamefont {Zhang}}, \bibinfo {author} {\bibfnamefont {D.~J.}\ \bibnamefont {Kelly}}, \bibinfo {author} {\bibfnamefont {R.~T.}\ \bibnamefont {Dahm}}, \bibinfo {author} {\bibfnamefont {C.~K.}\ \bibnamefont {Kirchert}}, \bibinfo {author} {\bibfnamefont {T.~E.~L.}\ \bibnamefont {Cozannet}}, \bibinfo {author} {\bibfnamefont {F.}~\bibnamefont {Trier}}, \emph {et~al.},\ }\bibfield  {title} {\bibinfo {title} {Stacking and twisting of freestanding complex oxide thin films},\ }\href@noop {} {\bibfield  {journal} {\bibinfo  {journal} {Adv. Mater.}\ }\textbf {\bibinfo {volume} {34}},\ \bibinfo {pages} {2203187} (\bibinfo {year} {2022})}\BibitemShut {NoStop}%
\bibitem [{\citenamefont {S{\'a}nchez-Santolino}\ \emph {et~al.}(2024)\citenamefont {S{\'a}nchez-Santolino}, \citenamefont {Rouco}, \citenamefont {Puebla}, \citenamefont {Aramberri}, \citenamefont {Zamora}, \citenamefont {Cabero}, \citenamefont {Cuellar}, \citenamefont {Munuera}, \citenamefont {Mompean}, \citenamefont {Garcia-Hernandez} \emph {et~al.}}]{sanchez20242d}%
  \BibitemOpen
  \bibfield  {author} {\bibinfo {author} {\bibfnamefont {G.}~\bibnamefont {S{\'a}nchez-Santolino}}, \bibinfo {author} {\bibfnamefont {V.}~\bibnamefont {Rouco}}, \bibinfo {author} {\bibfnamefont {S.}~\bibnamefont {Puebla}}, \bibinfo {author} {\bibfnamefont {H.}~\bibnamefont {Aramberri}}, \bibinfo {author} {\bibfnamefont {V.}~\bibnamefont {Zamora}}, \bibinfo {author} {\bibfnamefont {M.}~\bibnamefont {Cabero}}, \bibinfo {author} {\bibfnamefont {F.}~\bibnamefont {Cuellar}}, \bibinfo {author} {\bibfnamefont {C.}~\bibnamefont {Munuera}}, \bibinfo {author} {\bibfnamefont {F.}~\bibnamefont {Mompean}}, \bibinfo {author} {\bibfnamefont {M.}~\bibnamefont {Garcia-Hernandez}}, \emph {et~al.},\ }\bibfield  {title} {\bibinfo {title} {{A 2D ferroelectric vortex pattern in twisted BaTiO$_3$ freestanding layers}},\ }\href@noop {} {\bibfield  {journal} {\bibinfo  {journal} {Nature}\ }\textbf {\bibinfo {volume} {626}},\ \bibinfo {pages} {529} (\bibinfo {year} {2024})}\BibitemShut {NoStop}%
\bibitem [{\citenamefont {Kohn}\ and\ \citenamefont {Sham}(1965)}]{Kohn1965}%
  \BibitemOpen
  \bibfield  {author} {\bibinfo {author} {\bibfnamefont {W.}~\bibnamefont {Kohn}}\ and\ \bibinfo {author} {\bibfnamefont {L.~J.}\ \bibnamefont {Sham}},\ }\bibfield  {title} {\bibinfo {title} {{Self-Consistent Equations Including Exchange and Correlation Effects}},\ }\href {https://doi.org/10.1103/PhysRev.140.A1133} {\bibfield  {journal} {\bibinfo  {journal} {Phys. Rev.}\ }\textbf {\bibinfo {volume} {140}},\ \bibinfo {pages} {A1133} (\bibinfo {year} {1965})}\BibitemShut {NoStop}%
\bibitem [{\citenamefont {Kresse}\ and\ \citenamefont {Furthm\"{u}ller}(1996)}]{Kresse1996}%
  \BibitemOpen
  \bibfield  {author} {\bibinfo {author} {\bibfnamefont {G.}~\bibnamefont {Kresse}}\ and\ \bibinfo {author} {\bibfnamefont {J.}~\bibnamefont {Furthm\"{u}ller}},\ }\bibfield  {title} {\bibinfo {title} {{Efficient Iterative Schemes for $ab$ $initio$ Total-Energy Calculations Using a Plane-Wave Basis Set}},\ }\href {https://doi.org/10.1103/PhysRevB.54.11169} {\bibfield  {journal} {\bibinfo  {journal} {Phys. Rev. B}\ }\textbf {\bibinfo {volume} {54}},\ \bibinfo {pages} {11169} (\bibinfo {year} {1996})}\BibitemShut {NoStop}%
\bibitem [{\citenamefont {Ricciardulli}\ \emph {et~al.}(2021)\citenamefont {Ricciardulli}, \citenamefont {Yang}, \citenamefont {Smet},\ and\ \citenamefont {Saliba}}]{ricciardulli2021emerging}%
  \BibitemOpen
  \bibfield  {author} {\bibinfo {author} {\bibfnamefont {A.~G.}\ \bibnamefont {Ricciardulli}}, \bibinfo {author} {\bibfnamefont {S.}~\bibnamefont {Yang}}, \bibinfo {author} {\bibfnamefont {J.~H.}\ \bibnamefont {Smet}},\ and\ \bibinfo {author} {\bibfnamefont {M.}~\bibnamefont {Saliba}},\ }\bibfield  {title} {\bibinfo {title} {Emerging perovskite monolayers},\ }\href@noop {} {\bibfield  {journal} {\bibinfo  {journal} {Nat. Mater.}\ }\textbf {\bibinfo {volume} {20}},\ \bibinfo {pages} {1325} (\bibinfo {year} {2021})}\BibitemShut {NoStop}%
\bibitem [{\citenamefont {Deleuze}\ \emph {et~al.}(2022)\citenamefont {Deleuze}, \citenamefont {Magnan}, \citenamefont {Barbier}, \citenamefont {Li}, \citenamefont {Verdini}, \citenamefont {Floreano}, \citenamefont {Domenichini},\ and\ \citenamefont {Dupont}}]{deleuze2022nature}%
  \BibitemOpen
  \bibfield  {author} {\bibinfo {author} {\bibfnamefont {P.-M.}\ \bibnamefont {Deleuze}}, \bibinfo {author} {\bibfnamefont {H.}~\bibnamefont {Magnan}}, \bibinfo {author} {\bibfnamefont {A.}~\bibnamefont {Barbier}}, \bibinfo {author} {\bibfnamefont {Z.}~\bibnamefont {Li}}, \bibinfo {author} {\bibfnamefont {A.}~\bibnamefont {Verdini}}, \bibinfo {author} {\bibfnamefont {L.}~\bibnamefont {Floreano}}, \bibinfo {author} {\bibfnamefont {B.}~\bibnamefont {Domenichini}},\ and\ \bibinfo {author} {\bibfnamefont {C.}~\bibnamefont {Dupont}},\ }\bibfield  {title} {\bibinfo {title} {{Nature of the Ba 4d Splitting in BaTiO$_3$ Unraveled by a Combined Experimental and Theoretical Study}},\ }\href@noop {} {\bibfield  {journal} {\bibinfo  {journal} {J. Phys. Chem. C}\ }\textbf {\bibinfo {volume} {126}},\ \bibinfo {pages} {15899} (\bibinfo {year} {2022})}\BibitemShut {NoStop}%
\bibitem [{\citenamefont {Zhou}\ \emph {et~al.}(2015)\citenamefont {Zhou}, \citenamefont {Han}, \citenamefont {Dai}, \citenamefont {Sun},\ and\ \citenamefont {Srolovitz}}]{zhou2015van}%
  \BibitemOpen
  \bibfield  {author} {\bibinfo {author} {\bibfnamefont {S.}~\bibnamefont {Zhou}}, \bibinfo {author} {\bibfnamefont {J.}~\bibnamefont {Han}}, \bibinfo {author} {\bibfnamefont {S.}~\bibnamefont {Dai}}, \bibinfo {author} {\bibfnamefont {J.}~\bibnamefont {Sun}},\ and\ \bibinfo {author} {\bibfnamefont {D.~J.}\ \bibnamefont {Srolovitz}},\ }\bibfield  {title} {\bibinfo {title} {van der waals bilayer energetics: Generalized stacking-fault energy of graphene, boron nitride, and graphene/boron nitride bilayers},\ }\href@noop {} {\bibfield  {journal} {\bibinfo  {journal} {Phys. Rev. B}\ }\textbf {\bibinfo {volume} {92}},\ \bibinfo {pages} {155438} (\bibinfo {year} {2015})}\BibitemShut {NoStop}%
\bibitem [{\citenamefont {Carr}\ \emph {et~al.}(2018{\natexlab{a}})\citenamefont {Carr}, \citenamefont {Massatt}, \citenamefont {Torrisi}, \citenamefont {Cazeaux}, \citenamefont {Luskin},\ and\ \citenamefont {Kaxiras}}]{carr2018relaxation}%
  \BibitemOpen
  \bibfield  {author} {\bibinfo {author} {\bibfnamefont {S.}~\bibnamefont {Carr}}, \bibinfo {author} {\bibfnamefont {D.}~\bibnamefont {Massatt}}, \bibinfo {author} {\bibfnamefont {S.~B.}\ \bibnamefont {Torrisi}}, \bibinfo {author} {\bibfnamefont {P.}~\bibnamefont {Cazeaux}}, \bibinfo {author} {\bibfnamefont {M.}~\bibnamefont {Luskin}},\ and\ \bibinfo {author} {\bibfnamefont {E.}~\bibnamefont {Kaxiras}},\ }\bibfield  {title} {\bibinfo {title} {Relaxation and domain formation in incommensurate two-dimensional heterostructures},\ }\href@noop {} {\bibfield  {journal} {\bibinfo  {journal} {Phys. Rev. B}\ }\textbf {\bibinfo {volume} {98}},\ \bibinfo {pages} {224102} (\bibinfo {year} {2018}{\natexlab{a}})}\BibitemShut {NoStop}%
\bibitem [{\citenamefont {{\"O}z{\c{c}}elik}\ \emph {et~al.}(2018)\citenamefont {{\"O}z{\c{c}}elik}, \citenamefont {Fathi}, \citenamefont {Azadani},\ and\ \citenamefont {Low}}]{ozccelik2018tin}%
  \BibitemOpen
  \bibfield  {author} {\bibinfo {author} {\bibfnamefont {V.~O.}\ \bibnamefont {{\"O}z{\c{c}}elik}}, \bibinfo {author} {\bibfnamefont {M.}~\bibnamefont {Fathi}}, \bibinfo {author} {\bibfnamefont {J.~G.}\ \bibnamefont {Azadani}},\ and\ \bibinfo {author} {\bibfnamefont {T.}~\bibnamefont {Low}},\ }\bibfield  {title} {\bibinfo {title} {Tin monochalcogenide heterostructures as mechanically rigid infrared band gap semiconductors},\ }\href@noop {} {\bibfield  {journal} {\bibinfo  {journal} {Phys. Rev. Mater.}\ }\textbf {\bibinfo {volume} {2}},\ \bibinfo {pages} {051003} (\bibinfo {year} {2018})}\BibitemShut {NoStop}%
\bibitem [{\citenamefont {Cantele}\ \emph {et~al.}(2020)\citenamefont {Cantele}, \citenamefont {Alfe}, \citenamefont {Conte}, \citenamefont {Cataudella}, \citenamefont {Ninno},\ and\ \citenamefont {Lucignano}}]{cantele2020structural}%
  \BibitemOpen
  \bibfield  {author} {\bibinfo {author} {\bibfnamefont {G.}~\bibnamefont {Cantele}}, \bibinfo {author} {\bibfnamefont {D.}~\bibnamefont {Alfe}}, \bibinfo {author} {\bibfnamefont {F.}~\bibnamefont {Conte}}, \bibinfo {author} {\bibfnamefont {V.}~\bibnamefont {Cataudella}}, \bibinfo {author} {\bibfnamefont {D.}~\bibnamefont {Ninno}},\ and\ \bibinfo {author} {\bibfnamefont {P.}~\bibnamefont {Lucignano}},\ }\bibfield  {title} {\bibinfo {title} {Structural relaxation and low-energy properties of twisted bilayer graphene},\ }\href@noop {} {\bibfield  {journal} {\bibinfo  {journal} {Phys. Rev. Res.}\ }\textbf {\bibinfo {volume} {2}},\ \bibinfo {pages} {043127} (\bibinfo {year} {2020})}\BibitemShut {NoStop}%
\bibitem [{\citenamefont {Kim}\ \emph {et~al.}(2022{\natexlab{b}})\citenamefont {Kim}, \citenamefont {Ko}, \citenamefont {Jo}, \citenamefont {Kim}, \citenamefont {Yoo}, \citenamefont {Son},\ and\ \citenamefont {Cheong}}]{kim2022anomalous}%
  \BibitemOpen
  \bibfield  {author} {\bibinfo {author} {\bibfnamefont {J.}~\bibnamefont {Kim}}, \bibinfo {author} {\bibfnamefont {E.}~\bibnamefont {Ko}}, \bibinfo {author} {\bibfnamefont {J.}~\bibnamefont {Jo}}, \bibinfo {author} {\bibfnamefont {M.}~\bibnamefont {Kim}}, \bibinfo {author} {\bibfnamefont {H.}~\bibnamefont {Yoo}}, \bibinfo {author} {\bibfnamefont {Y.-W.}\ \bibnamefont {Son}},\ and\ \bibinfo {author} {\bibfnamefont {H.}~\bibnamefont {Cheong}},\ }\bibfield  {title} {\bibinfo {title} {Anomalous optical excitations from arrays of whirlpooled lattice distortions in moir{\'e} superlattices},\ }\href@noop {} {\bibfield  {journal} {\bibinfo  {journal} {Nat. Mater.}\ }\textbf {\bibinfo {volume} {21}},\ \bibinfo {pages} {890} (\bibinfo {year} {2022}{\natexlab{b}})}\BibitemShut {NoStop}%
\bibitem [{\citenamefont {Bennett}\ \emph {et~al.}(2023)\citenamefont {Bennett}, \citenamefont {Chaudhary}, \citenamefont {Slager}, \citenamefont {Bousquet},\ and\ \citenamefont {Ghosez}}]{bennett2023polar}%
  \BibitemOpen
  \bibfield  {author} {\bibinfo {author} {\bibfnamefont {D.}~\bibnamefont {Bennett}}, \bibinfo {author} {\bibfnamefont {G.}~\bibnamefont {Chaudhary}}, \bibinfo {author} {\bibfnamefont {R.-J.}\ \bibnamefont {Slager}}, \bibinfo {author} {\bibfnamefont {E.}~\bibnamefont {Bousquet}},\ and\ \bibinfo {author} {\bibfnamefont {P.}~\bibnamefont {Ghosez}},\ }\bibfield  {title} {\bibinfo {title} {Polar meron-antimeron networks in strained and twisted bilayers},\ }\href@noop {} {\bibfield  {journal} {\bibinfo  {journal} {Nat. Commun.}\ }\textbf {\bibinfo {volume} {14}},\ \bibinfo {pages} {1629} (\bibinfo {year} {2023})}\BibitemShut {NoStop}%
\bibitem [{\citenamefont {Junquera}\ and\ \citenamefont {Ghosez}(2003)}]{junquera2003critical}%
  \BibitemOpen
  \bibfield  {author} {\bibinfo {author} {\bibfnamefont {J.}~\bibnamefont {Junquera}}\ and\ \bibinfo {author} {\bibfnamefont {P.}~\bibnamefont {Ghosez}},\ }\bibfield  {title} {\bibinfo {title} {Critical thickness for ferroelectricity in perovskite ultrathin films},\ }\href@noop {} {\bibfield  {journal} {\bibinfo  {journal} {Nature}\ }\textbf {\bibinfo {volume} {422}},\ \bibinfo {pages} {506} (\bibinfo {year} {2003})}\BibitemShut {NoStop}%
\bibitem [{\citenamefont {Kim}\ \emph {et~al.}(2005)\citenamefont {Kim}, \citenamefont {Kim}, \citenamefont {Kim}, \citenamefont {Chang}, \citenamefont {Noh}, \citenamefont {Kong}, \citenamefont {Char}, \citenamefont {Park}, \citenamefont {Bu}, \citenamefont {Yoon} \emph {et~al.}}]{kim2005critical}%
  \BibitemOpen
  \bibfield  {author} {\bibinfo {author} {\bibfnamefont {Y.}~\bibnamefont {Kim}}, \bibinfo {author} {\bibfnamefont {D.}~\bibnamefont {Kim}}, \bibinfo {author} {\bibfnamefont {J.}~\bibnamefont {Kim}}, \bibinfo {author} {\bibfnamefont {Y.}~\bibnamefont {Chang}}, \bibinfo {author} {\bibfnamefont {T.}~\bibnamefont {Noh}}, \bibinfo {author} {\bibfnamefont {J.}~\bibnamefont {Kong}}, \bibinfo {author} {\bibfnamefont {K.}~\bibnamefont {Char}}, \bibinfo {author} {\bibfnamefont {Y.}~\bibnamefont {Park}}, \bibinfo {author} {\bibfnamefont {S.}~\bibnamefont {Bu}}, \bibinfo {author} {\bibfnamefont {J.-G.}\ \bibnamefont {Yoon}}, \emph {et~al.},\ }\bibfield  {title} {\bibinfo {title} {Critical thickness of ultrathin ferroelectric batio3 films},\ }\href@noop {} {\bibfield  {journal} {\bibinfo  {journal} {Appl. Phys. Lett.}\ }\textbf {\bibinfo {volume} {86}} (\bibinfo {year} {2005})}\BibitemShut {NoStop}%
\bibitem [{\citenamefont {Kim}\ \emph {et~al.}(2024)\citenamefont {Kim}, \citenamefont {Dominguez}, \citenamefont {Mayorga-Luna}, \citenamefont {Ye}, \citenamefont {Embley}, \citenamefont {Tan}, \citenamefont {Ni}, \citenamefont {Liu}, \citenamefont {Ford}, \citenamefont {Gao} \emph {et~al.}}]{kim2024electrostatic}%
  \BibitemOpen
  \bibfield  {author} {\bibinfo {author} {\bibfnamefont {D.~S.}\ \bibnamefont {Kim}}, \bibinfo {author} {\bibfnamefont {R.~C.}\ \bibnamefont {Dominguez}}, \bibinfo {author} {\bibfnamefont {R.}~\bibnamefont {Mayorga-Luna}}, \bibinfo {author} {\bibfnamefont {D.}~\bibnamefont {Ye}}, \bibinfo {author} {\bibfnamefont {J.}~\bibnamefont {Embley}}, \bibinfo {author} {\bibfnamefont {T.}~\bibnamefont {Tan}}, \bibinfo {author} {\bibfnamefont {Y.}~\bibnamefont {Ni}}, \bibinfo {author} {\bibfnamefont {Z.}~\bibnamefont {Liu}}, \bibinfo {author} {\bibfnamefont {M.}~\bibnamefont {Ford}}, \bibinfo {author} {\bibfnamefont {F.~Y.}\ \bibnamefont {Gao}}, \emph {et~al.},\ }\bibfield  {title} {\bibinfo {title} {Electrostatic moir{\'e} potential from twisted hexagonal boron nitride layers},\ }\href@noop {} {\bibfield  {journal} {\bibinfo  {journal} {Nat. Mater.}\ }\textbf {\bibinfo {volume} {23}},\ \bibinfo {pages} {65} (\bibinfo {year} {2024})}\BibitemShut {NoStop}%
\bibitem [{\citenamefont {Carr}\ \emph {et~al.}(2018{\natexlab{b}})\citenamefont {Carr}, \citenamefont {Fang}, \citenamefont {Jarillo-Herrero},\ and\ \citenamefont {Kaxiras}}]{carr2018pressure}%
  \BibitemOpen
  \bibfield  {author} {\bibinfo {author} {\bibfnamefont {S.}~\bibnamefont {Carr}}, \bibinfo {author} {\bibfnamefont {S.}~\bibnamefont {Fang}}, \bibinfo {author} {\bibfnamefont {P.}~\bibnamefont {Jarillo-Herrero}},\ and\ \bibinfo {author} {\bibfnamefont {E.}~\bibnamefont {Kaxiras}},\ }\bibfield  {title} {\bibinfo {title} {Pressure dependence of the magic twist angle in graphene superlattices},\ }\href@noop {} {\bibfield  {journal} {\bibinfo  {journal} {Phys. Rev. B}\ }\textbf {\bibinfo {volume} {98}},\ \bibinfo {pages} {085144} (\bibinfo {year} {2018}{\natexlab{b}})}\BibitemShut {NoStop}%
\bibitem [{\citenamefont {Xian}\ \emph {et~al.}(2021)\citenamefont {Xian}, \citenamefont {Claassen}, \citenamefont {Kiese}, \citenamefont {Scherer}, \citenamefont {Trebst}, \citenamefont {Kennes},\ and\ \citenamefont {Rubio}}]{xian2021realization}%
  \BibitemOpen
  \bibfield  {author} {\bibinfo {author} {\bibfnamefont {L.}~\bibnamefont {Xian}}, \bibinfo {author} {\bibfnamefont {M.}~\bibnamefont {Claassen}}, \bibinfo {author} {\bibfnamefont {D.}~\bibnamefont {Kiese}}, \bibinfo {author} {\bibfnamefont {M.~M.}\ \bibnamefont {Scherer}}, \bibinfo {author} {\bibfnamefont {S.}~\bibnamefont {Trebst}}, \bibinfo {author} {\bibfnamefont {D.~M.}\ \bibnamefont {Kennes}},\ and\ \bibinfo {author} {\bibfnamefont {A.}~\bibnamefont {Rubio}},\ }\bibfield  {title} {\bibinfo {title} {Realization of nearly dispersionless bands with strong orbital anisotropy from destructive interference in twisted bilayer mos2},\ }\href@noop {} {\bibfield  {journal} {\bibinfo  {journal} {Nat. Commun.}\ }\textbf {\bibinfo {volume} {12}},\ \bibinfo {pages} {5644} (\bibinfo {year} {2021})}\BibitemShut {NoStop}%
\bibitem [{\citenamefont {Tao}\ \emph {et~al.}(2022)\citenamefont {Tao}, \citenamefont {Zhang}, \citenamefont {Zhu}, \citenamefont {He}, \citenamefont {Yang}, \citenamefont {Lu},\ and\ \citenamefont {Wei}}]{tao2022designing}%
  \BibitemOpen
  \bibfield  {author} {\bibinfo {author} {\bibfnamefont {S.}~\bibnamefont {Tao}}, \bibinfo {author} {\bibfnamefont {X.}~\bibnamefont {Zhang}}, \bibinfo {author} {\bibfnamefont {J.}~\bibnamefont {Zhu}}, \bibinfo {author} {\bibfnamefont {P.}~\bibnamefont {He}}, \bibinfo {author} {\bibfnamefont {S.~A.}\ \bibnamefont {Yang}}, \bibinfo {author} {\bibfnamefont {Y.}~\bibnamefont {Lu}},\ and\ \bibinfo {author} {\bibfnamefont {S.-H.}\ \bibnamefont {Wei}},\ }\bibfield  {title} {\bibinfo {title} {Designing ultra-flat bands in twisted bilayer materials at large twist angles: theory and application to two-dimensional indium selenide},\ }\href@noop {} {\bibfield  {journal} {\bibinfo  {journal} {J. Am. Chem. Soc.}\ }\textbf {\bibinfo {volume} {144}},\ \bibinfo {pages} {3949} (\bibinfo {year} {2022})}\BibitemShut {NoStop}%
\bibitem [{\citenamefont {Jiang}\ \emph {et~al.}(2019)\citenamefont {Jiang}, \citenamefont {Kang}, \citenamefont {Huang}, \citenamefont {Xu}, \citenamefont {Low},\ and\ \citenamefont {Liu}}]{jiang2019topological}%
  \BibitemOpen
  \bibfield  {author} {\bibinfo {author} {\bibfnamefont {W.}~\bibnamefont {Jiang}}, \bibinfo {author} {\bibfnamefont {M.}~\bibnamefont {Kang}}, \bibinfo {author} {\bibfnamefont {H.}~\bibnamefont {Huang}}, \bibinfo {author} {\bibfnamefont {H.}~\bibnamefont {Xu}}, \bibinfo {author} {\bibfnamefont {T.}~\bibnamefont {Low}},\ and\ \bibinfo {author} {\bibfnamefont {F.}~\bibnamefont {Liu}},\ }\bibfield  {title} {\bibinfo {title} {Topological band evolution between lieb and kagome lattices},\ }\href@noop {} {\bibfield  {journal} {\bibinfo  {journal} {Phys. Rev. B}\ }\textbf {\bibinfo {volume} {99}},\ \bibinfo {pages} {125131} (\bibinfo {year} {2019})}\BibitemShut {NoStop}%
\bibitem [{\citenamefont {Jiang}\ \emph {et~al.}(2020)\citenamefont {Jiang}, \citenamefont {Zhang}, \citenamefont {Wang}, \citenamefont {Liu},\ and\ \citenamefont {Low}}]{jiang2020topological}%
  \BibitemOpen
  \bibfield  {author} {\bibinfo {author} {\bibfnamefont {W.}~\bibnamefont {Jiang}}, \bibinfo {author} {\bibfnamefont {S.}~\bibnamefont {Zhang}}, \bibinfo {author} {\bibfnamefont {Z.}~\bibnamefont {Wang}}, \bibinfo {author} {\bibfnamefont {F.}~\bibnamefont {Liu}},\ and\ \bibinfo {author} {\bibfnamefont {T.}~\bibnamefont {Low}},\ }\bibfield  {title} {\bibinfo {title} {Topological band engineering of lieb lattice in phthalocyanine-based metal--organic frameworks},\ }\href@noop {} {\bibfield  {journal} {\bibinfo  {journal} {Nano Lett.}\ }\textbf {\bibinfo {volume} {20}},\ \bibinfo {pages} {1959} (\bibinfo {year} {2020})}\BibitemShut {NoStop}%
\bibitem [{\citenamefont {Cui}\ \emph {et~al.}(2020)\citenamefont {Cui}, \citenamefont {Zheng}, \citenamefont {Wang}, \citenamefont {Liu}, \citenamefont {Xie},\ and\ \citenamefont {Huang}}]{cui2020realization}%
  \BibitemOpen
  \bibfield  {author} {\bibinfo {author} {\bibfnamefont {B.}~\bibnamefont {Cui}}, \bibinfo {author} {\bibfnamefont {X.}~\bibnamefont {Zheng}}, \bibinfo {author} {\bibfnamefont {J.}~\bibnamefont {Wang}}, \bibinfo {author} {\bibfnamefont {D.}~\bibnamefont {Liu}}, \bibinfo {author} {\bibfnamefont {S.}~\bibnamefont {Xie}},\ and\ \bibinfo {author} {\bibfnamefont {B.}~\bibnamefont {Huang}},\ }\bibfield  {title} {\bibinfo {title} {Realization of lieb lattice in covalent-organic frameworks with tunable topology and magnetism},\ }\href@noop {} {\bibfield  {journal} {\bibinfo  {journal} {Nat. Commun.}\ }\textbf {\bibinfo {volume} {11}},\ \bibinfo {pages} {66} (\bibinfo {year} {2020})}\BibitemShut {NoStop}%
\bibitem [{\citenamefont {Bl\"{o}chl}(1994)}]{Blochl1994}%
  \BibitemOpen
  \bibfield  {author} {\bibinfo {author} {\bibfnamefont {P.~E.}\ \bibnamefont {Bl\"{o}chl}},\ }\bibfield  {title} {\bibinfo {title} {{Projector Augmented-Wave Method}},\ }\href@noop {} {\bibfield  {journal} {\bibinfo  {journal} {Phys. Rev. B}\ }\textbf {\bibinfo {volume} {50}},\ \bibinfo {pages} {17953} (\bibinfo {year} {1994})}\BibitemShut {NoStop}%
\bibitem [{\citenamefont {Kresse}\ and\ \citenamefont {Joubert}(1999)}]{Kresse1999}%
  \BibitemOpen
  \bibfield  {author} {\bibinfo {author} {\bibfnamefont {G.}~\bibnamefont {Kresse}}\ and\ \bibinfo {author} {\bibfnamefont {D.}~\bibnamefont {Joubert}},\ }\bibfield  {title} {\bibinfo {title} {{From Ultrasoft Pseudopotentials to the Projector Augmented-Wave Method}},\ }\href {https://doi.org/10.1103/PhysRevB.59.1758} {\bibfield  {journal} {\bibinfo  {journal} {Phys. Rev. B}\ }\textbf {\bibinfo {volume} {59}},\ \bibinfo {pages} {1758} (\bibinfo {year} {1999})}\BibitemShut {NoStop}%
\bibitem [{\citenamefont {Perdew}\ \emph {et~al.}(1996)\citenamefont {Perdew}, \citenamefont {Burke},\ and\ \citenamefont {Ernzerhof}}]{Perdew1996}%
  \BibitemOpen
  \bibfield  {author} {\bibinfo {author} {\bibfnamefont {J.~P.}\ \bibnamefont {Perdew}}, \bibinfo {author} {\bibfnamefont {K.}~\bibnamefont {Burke}},\ and\ \bibinfo {author} {\bibfnamefont {M.}~\bibnamefont {Ernzerhof}},\ }\bibfield  {title} {\bibinfo {title} {{Generalized Gradient Approximation Made Simple}},\ }\href {https://doi.org/10.1103/PhysRevLett.77.3865} {\bibfield  {journal} {\bibinfo  {journal} {Phys. Rev. Lett.}\ }\textbf {\bibinfo {volume} {77}},\ \bibinfo {pages} {3865} (\bibinfo {year} {1996})}\BibitemShut {NoStop}%
\bibitem [{\citenamefont {Grimme}\ \emph {et~al.}(2010)\citenamefont {Grimme}, \citenamefont {Antony}, \citenamefont {Ehrlich},\ and\ \citenamefont {Krieg}}]{grimme-d3}%
  \BibitemOpen
  \bibfield  {author} {\bibinfo {author} {\bibfnamefont {S.}~\bibnamefont {Grimme}}, \bibinfo {author} {\bibfnamefont {J.}~\bibnamefont {Antony}}, \bibinfo {author} {\bibfnamefont {S.}~\bibnamefont {Ehrlich}},\ and\ \bibinfo {author} {\bibfnamefont {H.}~\bibnamefont {Krieg}},\ }\bibfield  {title} {\bibinfo {title} {{A consistent and accurate ab initio parametrization of density functional dispersion correction (DFT-D) for the 94 elements H-Pu}},\ }\href@noop {} {\bibfield  {journal} {\bibinfo  {journal} {J. Chem. Phys.}\ }\textbf {\bibinfo {volume} {132}} (\bibinfo {year} {2010})}\BibitemShut {NoStop}%
\end{thebibliography}

%

\end{document}